\documentclass[preprint,12pt,authoryear]{elsarticle}

\usepackage{amssymb}
\usepackage{amsmath}

\usepackage{lineno}

\usepackage{booktabs}
\usepackage{caption}
\usepackage{multirow}
\usepackage{xurl}
\usepackage{hyperref}
\usepackage[capitalize]{cleveref}
\usepackage{todonotes}

\usepackage{makecell}

\newcommand{\coarsenet}{\emph{CoarseSoundNet}}
\newcommand{\edansa}{\emph{Edansa-2019}}
\newcommand{\besound}{\emph{BEsound}}

\newcommand{\akwamo}{\emph{BrPAM}} 
\newcommand{\synthdata}{\emph{PublicMix}} 
\newcommand{\htsforest}{\emph{HTS-Forest}} 
\newcommand{\beambient}{\emph{BE-Ambient}} 

\newcommand{\eg}{e.\,g.}
\newcommand{\ie}{i.\,e.}
\newcommand{\wrt}{w.\,r.\,t.\ }

\usepackage[acronym, shortcuts, nohypertypes={acronym}]{glossaries}
\newacronym{ACC}{ACC}{accuracy}
\newacronym{ACI}{ACI}{acoustic complexity index}
\newacronym{ADI}{ADI}{acoustic diversity index}
\newacronym{aROI}{aROI}{area of region of interest}
\newacronym{BE}{BE}{biodiversity exploratories}
\newacronym{CBT}{CBT}{count-based thresholding}
\newacronym{CI}{CI}{confidence interval}
\newacronym{CNN}{CNN}{convolutional neural network}
\newacronym{CST}{CST}{class-specific thresholds}
\newacronym{DCASE}{DCASE}{detection and classification of acoustic scenes and events}
\newacronym{DL}{DL}{deep learning}
\newacronym{HTS}{HTS}{HearTheSpecies}
\newacronym{LLM}{LLM}{large language model}
\newacronym{MIT}{MIT}{mean inference time}
\newacronym{ML}{ML}{machine learning}
\newacronym{MS}{MS}{model soup}
\newacronym{NDSI}{NDSI}{normalised difference soundscape index}
\newacronym{OOD}{OOD}{out-of-domain}
\newacronym{PAM}{PAM}{passive acoustic monitoring}
\newacronym{PAR}{PAR}{passive acoustic recorder}
\newacronym{PDA}{PDA}{proportional duration annotation}
\newacronym{PR}{PR}{precision-recall}
\newacronym{ROC}{ROC}{receiver operating characteristic}
\newacronym{SM}{SM}{song meter}
\newacronym{TBA}{TBA}{time-based adapted annotations}

\usepackage[normalem]{ulem}  

\definecolor{tcadd}{rgb}{0.0,0.6,0.0}
\definecolor{tcdel}{rgb}{0.1,0.1,0.8}
\definecolor{tcchg}{rgb}{0.33,0.11,0.46}

\journal{Ecological Informatics}

\begin{document}

\begin{frontmatter}



\title{CoarseSoundNet: Building a reliable model \\for ecological soundscape analysis} 

\author[1,3]{Alexander Gebhard\corref{cor1}}
\author[1,3]{Andreas Triantafyllopoulos}
\author[2]{Dominik Arend}
\author[2]{Sandra Müller}
\author[2]{Svenja Schmidt}
\author[2]{Michael Scherer-Lorenzen}
\author[1,3,4]{Björn W. Schuller}
\cortext[cor1]{Corresponding author, email address: alexander.gebhard@tum.de}
\affiliation[1]{organization={TUM University Hospital, CHI -- Chair of Health Informatics},
            addressline={Ismaninger Str. 22}, 
            city={Munich},
            postcode={81675}, 
            state={Bavaria},
            country={Germany}}

\affiliation[2]{organization={University of Freiburg, Faculty of Biology, Geobotany},
            addressline={Schaenzlestr. 1},
            city={Freiburg},
            postcode={79104},
            state={Baden-Württemberg},
            country={Germany}}

\affiliation[3]{organization={MCML -- Munich Center for Machine Learning},
            city={Munich},
            state={Bavaria},
            country={Germany}}

\affiliation[4]{organization={Imperial College London, GLAM -- Group on Language, Audio, \& Music},
            city={London},
            country={UK}}



\begin{abstract}
A soundscape is composed of three types of sound: biophony (sounds made by animals), geophony (natural abiotic sounds) and anthropophony (sounds made by humans).  A key research question in the field of soundscape ecology is how these components interact with each other, specifically how biophony responds to geophony and anthropophony. Nevertheless, as of today, there are not many analytical instruments that enable the distinct quantification of these elements. Recent \ac{ML} approaches aim to support automated analysis but often rely on task-specific or clean data, limiting generalisation to noisy \ac{PAM} recordings.
This study presents a reproducible framework to build \ac{ML} models for coarse soundscape classification and introduces {\coarsenet}, a deep learning model trained to distinguish biophony, geophony, and anthropophony under realistic terrestrial \ac{PAM} conditions. We systematically investigate model architectures, the influence of an additional training class, data composition, and evaluation strategies. 
Our findings suggest that model performance improves with additional \ac{PAM} data, especially when similar to the target domain, and by introducing an explicit silence class during training. 
Post-hoc target-domain calibration employing class-specific decision thresholds further enhance performance, while incorporating minimum annotation durations yields higher scores for anthropophony and geophony.
Error analyses exhibit challenges for anthropophony due to masking effects and confusions for silence and insect sounds for geophony and biophony.
Finally, we conduct an ecological case study showing that pre-filtering recordings with {\coarsenet} yields acoustic index trends broadly similar to ground-truth filtering, although the resulting improvements in their relationships with avian $\alpha$-diversity were limited.

\end{abstract}



\begin{keyword}
ecoacoustics \sep soundscape ecology \sep machine learning \sep deep learning \sep computer audition



\end{keyword}

\end{frontmatter}


\section{Introduction}
\label{sec:introduction}

Since the founding of \textit{soundscape ecology} as an interdisciplinary field \citep{Pijanowski11-SET} the question of how the different components of a soundscape interact on a landscape scale has been a key research question to this field. Abiotic sounds, especially sounds from engines (technophony) but also natural abiotic sounds from wind and rain, have profound impacts on wildlife and their communication \citep{Francis23-BAI}. 
Current research investigating the effect of traffic and engine noises on wildlife estimates noise impacts by approximating distance to the nearest road or settlements and noise modelling approaches \citep{Cooke20-RAA,Doser20-CFR,Ghadirian19-IND,Konstantopoulos20-ASE}. Due to a lack of tools, few studies to date, utilise the power of \ac{PAM} schemes combined with \ac{ML}-based models to classify the different soundscape components, to measure prevalence of geophony and anthropophony in-situ and their direct effects on biophony. The few studies that do so show that anthropogenic sounds (airplanes in that case) are of higher prevalence than conventional noise mapping approaches would reveal \citep{Grinfeder22-SDO}. 

This gap has become particularly relevant with the rapid growth of \ac{PAM}, supported by affordable recording devices such as AudioMoths \citep{Hill18-AEO} and advances in big data, \ac{ML}, and \ac{DL} \citep{Stowell22-CBW,Triantafyllopoulos25-CAF}. These developments enable the automated analysis of large-scale acoustic datasets, which would otherwise require extensive manual effort. As a result, \ac{ML}-based methods have become central to modern bioacoustic and ecoacoustic research \citep{Stowell22-CBW}.

Following this trend, \ac{ML} models are now widely deployed in bioacoustics, particularly for species detection and classification. For birds, models, such as BirdNET \citep{Kahl21-BAD}, Perch \citep{Hamer23-BAG}, and systems developed in the \ac{DCASE} \citep{Stowell18-AAD} and BirdCLEF \citep{Kahl20-OOB,Lasseck19-BSI} challenges, have become central tools for large-scale monitoring and annotation. Similar approaches have been extended for bats \citep{Mac18-BDL,Tabak22-ACO,Kobayashi21-DOA,Triantafyllopoulos24-AAA}, orcas \citep{Bergler19-OAA,Bergler21-OAA}, and other taxa \citep{Himawan18-DLT,Lebien20-APF,Yin21-ALD,Dufourq21-ADO,Romero21-UDF,Jung21-DLC,Faiss25-IAO}. Recent work has further expanded these efforts to rare species with little data and low-resource settings, including few-shot and zero-shot approaches \citep{Kahl22-OOB,Morfi21-FBE,Moummad24-SLF,Gebhard24-EMI}, as well as foundation models, such as AVES and BirdAVES \citep{Hagiwara23-AAV}, and \acp{LLM}-based approaches to further improve performance, as in the case of Nature-LM \citep{Robinson24-NAA}.

However, even though these models enable species-specific analyses, they capture only a part of the bigger picture constituted by \textit{ecological soundscapes}. 
While there have been several definitions of ``soundscapes'' over the years \citep{Southworth67-TSE,Raimbault05-USE,Farina14-SEP,Pijanowski11-SET}, we opt for the one given by \citet{Pijanowski11-SET}, who shaped the term \textit{soundscape ecology} by drawing parallels to ecological landscapes. They define a soundscape as ``the collection of sounds that emanate from landscapes'' \citep{Pijanowski11-SET}, thus covering the relations and interactions among the three main components \textit{anthropophony} (human-produced sounds), \textit{biophony} (sounds by animals), and \textit{geophony} (natural abiotic sounds like wind or rain).

While \ac{ML}-models can detect and classify species, their presence, and vocal behaviour, classifying all soundscape components also enables a better understanding of how the acoustic habitat - including geophony and anthropophony - influences vocal behaviour and acoustic community composition \citep{Mullet17-TAH}. 
Human land use, land cover change, exploitation, biodiversity changes, and climate change all shape the composition and diversity of soundscapes, with impacts on ecological processes like communication and information exchanges, which in turn affect species composition, species behaviour, but also human well being and recreational values of landscapes \citep{Dumyahn2011-SCO,Mullet17-TAH,Pijanowski11-WIS}.
Moreover, the presence of non-biophonic sounds can affect the reliability of acoustic indices as well as the detection range and precision of \ac{ML}-based species classification models. In order to standardise \ac{PAM}-based species monitoring schemes, it is essential to identify recordings with geophony, including wind, and not only rain \citep{Metcalf20-HAR}.

In this context, we situate our work within the \ac{BE} (DFG Priority Programme 1374), a large-scale biodiversity research platform investigating the effects of land use on biodiversity and ecosystem processes \citep{Fischer10-ILA}.
In particular, we provide a comprehensive analysis of different design choices for creating a soundscape model, resulting in the creation and release of \emph{CoarseSoundNet}, a publicly-available model that will hopefully facilitate more robust analyses of soundscape data. 
In this study, we consider reliability in terms of robust model performance across different \ac{PAM} conditions and the ability to further calibrate the model to a specific target domain and task to achieve strong performance under the intended target conditions.

Specifically, we address the following research questions (RQs):
\textbf{RQ1:} Which model architectures and training strategies are most suitable for coarse soundscape classification, and how well do they transfer across \ac{PAM} domains?
\textbf{RQ2:} How do the composition and domain similarity of training data, as well as target-domain calibration and task-specific evaluation strategies, affect model performance?
\textbf{RQ3:} To what extent can coarse soundscape classification support preprocessing and complementary analyses in ecoacoustic applications?
Our model, the corresponding code, and the configuration files are available on huggingface\footnote{\url{https://huggingface.co/HearTheSpecies/CoarseSoundNet}}~\citep{Gebhard26-CAM} and github\footnote{\url{https://github.com/CHI-TUM/CoarseSoundNet}}.

\section{Related work}
\label{sec:related-work}

The traditional way of measuring soundscapes has been the utilisation of soundscape indices comprising, among others, \textit{acoustic indices} which aim to ``quantify the complexity, diversity, and/or breadth of sound sources in a soundscape'' \citep{Pijanowski24-POS}. Some commonly used indices are the \ac{ACI} \citep{Pieretti11-ANM}, the \ac{ADI} \citep{Villanueva11-APO}, or the \ac{NDSI} \citep{Kasten12-TRE}, which belong to the ``classic'' acoustic indices \citep{Pijanowski24-POS,Sueur14-AIF}. 
The \ac{ACI} calculates the relative intensity fluctuations between adjacent frequency bins over time and was intended to quantify biotic sounds~\citep{Pieretti11-ANM}. The \ac{ADI} computes the distribution of acoustic energy across frequency bands by using the Shannon index, indicating the diversity of acoustic activity~\citep{Villanueva11-APO,Pekin12-MAD}. The \ac{NDSI} was created to reflect the ratio of biophonic to anthropophonic sounds, and thus indicating the level of human disturbance in the soundscape~\citep{Kasten12-TRE}. \citet{Bradfer19-GFT} describe those indices and their purpose in more detail. 

Those indices have been utilised in several prior studies in order to measure environmental processes, soundscape components, and their interactions \citep{Alcocer22-AIA,Arend25-SEO,Bradfer23-UAI,Lai25-COS}.
The shortcomings, however, are that these indices have limited ecological specificity as they cannot distinguish between biophony, geophony, and anthropophony and rather represent indirect proxies than direct ecological measurements \citep{Alcocer22-AIA}. Furthermore, they are susceptible to non-biological noise like wind, rain, or traffic \citep{Fairbrass17-BOA}. \ac{ML} models can bridge those gaps for instance by providing class-specific information or a way of pre-filtering recordings based on soundscape categories, before applying these indices \citep{Arend25-SEO}.

Thus, recently, there is an increase in \ac{ML} models that are developed in order to attain more component-specific and thus in-depth and more robust insights on the respective classes and their relations;
This is done either by applying these models before calculating the acoustic indices, in order to filter certain recordings \citep{Arend25-SEO}, or by directly using the soundscape predictions of the model for the analysis \citep{Fairbrass19-CDL,Quinn22-SCW}.

Some of the \ac{ML}-based approaches only focus on certain classes, like CityNet, which focuses on anthropophony and biophony \citep{Fairbrass19-CDL} or \citet{Terranova24-WED}, who focus on wind, rain, and biophony, while others cover all three coarse soundscape classes \citep{Challeat24-ADO} or even more than the three main targets, by also comprising interference (\ie, electronic or physical microphone events), background sounds or silence \citep{Quinn22-SCW,Grinfeder22-SDO}, or having additional 
fine-grained annotations \citep{Coban22-ETE,Wang25-RDD,Jiang26-RNS}. 

However, most of these studies rely on study-specific datasets, collected from the same recording areas, using the same devices under similar recording conditions, which limits their ability to evaluate model performance under substantially different settings~\citep{Sethi23-LTT} as well as their broader applicability. Additionally, some studies have only limited or no \ac{PAM} data for model training, leading them to rely on opportunistically sourced datasets, such as AudioSet~\citep{Gemmeke17-ASA} or FreeSound~\citep{Fonseca17-FDA}, as some of their (additional) sources \citep{Challeat24-ADO,Quinn22-SCW}, without rigorously testing how the models perform in the \ac{PAM} domain.
These observations present a major gap in the current literature, given the substantial drop in performance when the training data differs substantially from the test data, a condition that is widely known in \ac{ML} literature as a \emph{domain mismatch}~\citep{Ben10-ATO}.
Furthermore, all of the above studies relied on variants of \acp{CNN}, thus potentially missing out on the advances of more contemporary, transformer-based architectures.

This manuscript addresses these gaps in prior work by investigating the factors which contribute to the in-domain and cross-domain performance of a \emph{coarse soundscape classification} model which aims to identify the presence of anthropophony, biophony, and geophony.
We begin by investigating the role of model architecture, (pre)training data, data augmentation, and task operationalisation in the success of trained models.
Through this process, we provide a recipe for building soundscape classification models and develop {\coarsenet}, a publicly-available model that can identify all three 
soundscape components in terrestrial \ac{PAM} data.
For all our experiments we leveraged the \textit{autrainer} library \citep{Rampp24-AAM}, a tool for deep learning training in computer audition tasks, to enable rapid and reproducible model training.

\section{Methodology}

This section describes our methodology:
\cref{ssec:data} presents the data used in our work, \cref{ssec:model-training} and \cref{ssec:silence} our in-domain experiments, and \cref{ssec:additional-data} our cross-domain evaluation.
\cref{ssec:evaluation-strategy} presents different operationalisations of a coarse classification task and how those impact performance. \cref{ssec:methodology-ecological-case-study} shows an application scenario of {\coarsenet}.

\subsection{Data}
\label{ssec:data}

\begin{figure}[t]
    \centering
    \includegraphics[width=\linewidth]{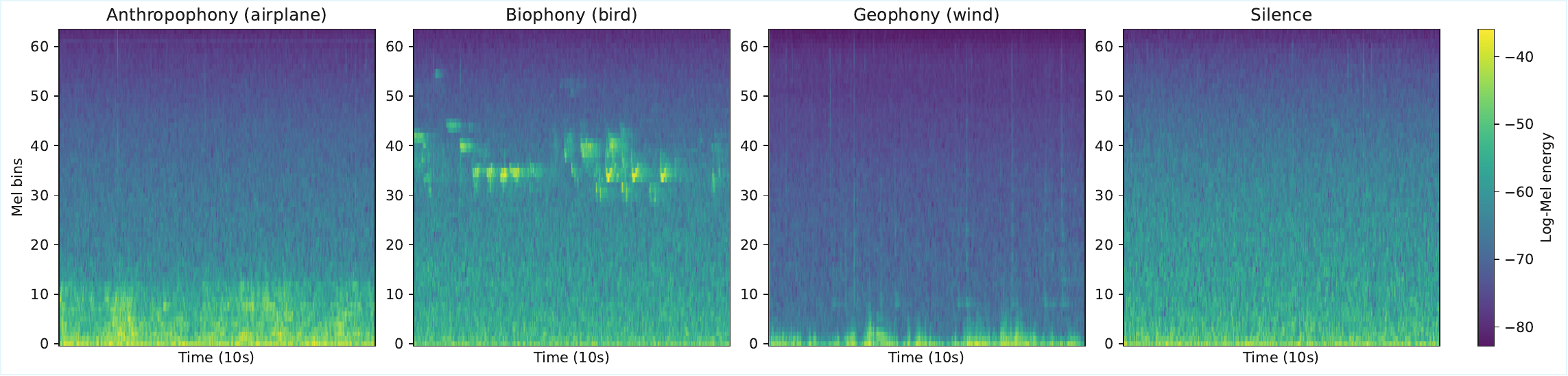}
    \caption{Example spectrograms for the four acoustic classes: anthropophony, biophony, geophony, and silence. Computed on {\besound} data.}
    \label{fig:spectrograms}
\end{figure}

Training our model required annotated audio data covering the three coarse soundscape categories: \textbf{Anthropophony}, \textbf{Biophony}, and \textbf{Geophony}. To this end, we employed a combination of publicly available datasets and private collections annotated by experts in biology and ecology.
The two primary datasets of our study were {\edansa} and {\besound}. The {\edansa} dataset, introduced by \citet{Coban22-ETE}, is publicly available, as well as {\besound} which was annotated specifically for this work by the third and fourth authors and their research group at the University of Freiburg. Specifically, {\besound} is a subset of the data which was collected during the {\besound} project, an affiliated project of the \ac{BE}~\citep{Mueller22-LIA,Mueller24-TDO}.

As {\besound} was collected in the \ac{BE} under recording conditions representative of the intended application of {\coarsenet}, it represents our main target domain. Our broader objective is to develop a model that transfers robustly across different terrestrial \ac{PAM} settings, while ensuring reliable performance under the recording conditions relevant to our applications within the BE. We therefore deliberately considered performance on both {\edansa} and {\besound} when making major model-development decisions to avoid selecting configurations that perform well on {\edansa} but transfer poorly to the target domain. Consequently, {\besound} should not be interpreted as a conventional held-out test set for the final model, but rather as a target-domain dataset that also informed model development.

In addition, we used four supplementary datasets: {\beambient}, {\htsforest}, and {\akwamo}, all collected and annotated by the University of Freiburg, as well as {\synthdata}, which consists of a curated mixture of different publicly available data sources. We made all of these datasets are publicly accessible via Zenodo (see \cref{sec:data-availability}).
The distribution of the coarse categories and their combinations within each dataset is summarised in \cref{tab:data-distribution}. The anthropophony samples predominantly comprise technophonic sounds, such as engine, traffic, aircraft, and machinery noise, but can also include other anthropogenic sounds, such as human speech and footsteps. In the following, we describe each dataset in more detail.

\begin{table}[t]
    \centering
    \resizebox{\textwidth}{!}{%
    \begin{tabular}{|l|rrrrrrrr|rrrr|r|r|}
        \hline
        Dataset & A    & B    & G   & S   & AB  & AG  & BG   & ABG & A$_t$ & B$_t$ & G$_t$ & S$_t$ & $\sum$ samples & $\sum$ hours\\
        \hline
        \edansa  & 2\,286 & 2\,509 & 357 & 446 & 313 & 345 & 3\,307 & 131 & 3\,076  & 6\,260  & 5\,214  & 1\,521  & 10\,771 & 29.9\\ 
        \hline
        \akwamo &   56 &  482 & 902  & 109  & 162  & 246  & 971  & 410 &  875  & 2\,025  & 2\,530  &  111  &  3\,341 & 9.3\\ 
        \beambient &  151 &   23 & 257  & 337  & 237  & 125  & 163  & 133 &  646  &  556  &  678  &  337  &  1\,426 & 2.0\\ 
        \htsforest &  293 &  138 & 283  & 148  &  98  & 111  &  35  &   8 &  510  &  279  &  437  &  148  &  1\,162 & 1.6 \\ 
        \synthdata & 8\,500 & 8\,500 & 8\,500 & 8\,500 & 8\,500 & 8\,498 & 8\,500 & 8\,499 & 33\,997 & 33\,999 & 33\,997 &  8\,500 &  67\,997 & 94.4 \\ 
        \hline
        \besound   &  281 &  835 & 483  & 851  & 542  & 347  &  447 & 461 & 1\,631  & 2\,285  & 1\,738  &  851  &  4\,247 & 70.8 \\ 
        \hline
    \end{tabular}
    }
\caption{Distribution of exclusive and combined acoustic classes across the datasets. 
$A = Anthropophony$, $B = Biophony$, $G = Geophony$, $S = Silence$. 
Combinations denote co-occurrence of classes. For instance, $AB$ comprises only samples that are annotated with both anthropophony and biophony, while $G$ refers to samples annotated only with geophony. 
The listed combinations are mutually exclusive but not exhaustive: co-occurrences involving silence and recordings without an active label are not shown separately.
A$_t$, B$_t$, G$_t$, S$_t$ denote the total counts of each class, independent of combinations. 
Class totals and total sample counts are calculated across all recordings, including those not represented by the listed combinations.
The total audio hours for each dataset are given in the last column.}
\label{tab:data-distribution}
\end{table}

\textbf{{\edansa}:} The {\edansa} dataset \citep{Coban22-ETE} is a publicly available multi-label ecoacoustic dataset of annotated soundscapes. 
It was collected on the Arctic North Slope of Alaska during summer 2019 using \ac{SM}-4 wildlife recorders (Wildlife Acoustics) and contains 28 hierarchical annotation tags, including the three target classes, as well as a ``Silence'' tag. The published recordings consist of 10\,s audio clips sampled at 48\,kHz.

One limitation of this dataset lies in its ambiguous definition of silence: several clips are simultaneously labelled with geophony (\eg, wind or rain) and silence. In our work, we re-defined silence to strictly denote the absence of all three main classes. 
Moreover, subsequent work by the same authors identified audio clips affected by ``clipping'', \ie, samples where the signal exceeded the recording device’s dynamic range, which may introduce distortions that could influence the reliability of the annotations and model training~\citep{Coban24-THR}.
Despite these drawbacks, {\edansa} remains the most valuable available resource for our task, as it is, to the best of our knowledge, the only publicly available soundscape dataset explicitly annotated for anthropophony, biophony, and geophony.
In this study, {\edansa} serves as the core dataset for training the model, choosing an appropriate training configuration, and selecting the final model architecture.

\textbf{{\besound}:}
{\besound} was annotated specifically for this study in order to assess model performance in the \ac{BE} regions in Germany. It comprises soundscapes recorded in forest 
and grassland habitat during spring and summer 2016. The audio clips are 60\,s long and were recorded with a prototype version of Soundscape Explorer Terrestrial (SET) by Lunilettronik at 48\,kHz.
As a domain distinct from {\edansa}, {\besound} was initially utilised to assess cross-domain transfer and additionally served as an important target-domain dataset that informed further model development.

\textbf{{\akwamo}:} These data are an excerpt of the {\akwamo} project (ID: 2221\-NR\-050A) and were annotated by the fourth author. Recordings were collected in forest habitat in the Britz region (Germany) deploying \ac{SM}-4 
wildlife recorders (Wildlife Acoustics). The provided clips are 10\,s long, sampled at 48\,kHz, and the annotations cover all three target classes as well as silence.

\textbf{{\beambient}:}
To increase coverage from regions similar to {\besound}, but a different time period (2015–2016), these data were annotated with the same coarse classes. The recordings are provided as 5\,s audio clips sampled at 48\,kHz and were recorded in forest habitats using a prototype version of Soundscape Explorer Terrestrial (SET) by Lunilettronik. 

\textbf{{\htsforest}:}
These data are a small portion of the recordings collected in the scope of the \ac{HTS} project\footnote{\url{https://gepris.dfg.de/gepris/projekt/512414116}} at the three \ac{BE} regions in Germany (in forest habitat). The audio was recorded during spring and summer 2023 with AudioMoths (versions $1.1.0$ and $1.2.0$) at 48\,kHz. The labelled clips are 5\,s in duration. All our target classes and silence are annotated. 

\textbf{{\synthdata}:}
In addition to the soundscape data sources above, we curated a mixed dataset by mixing audio from different public sources: Audio Set~\citep{Gemmeke17-ASA}, Orthoptera recordings from xeno-canto, FSD50K~\citep{Fonseca22-FAO}, IDMT-Traffic~\citep{Abeßer21-IAO}, MAVD \citep{Zinemanas19-MAD}, AeroSonicDB~\citep{Downward23-TAD}, WindNoiseDataset \citep{Yang22-WND}, and WindNet-data~\citep{Terranova24-WED}. All synthetic clips are mixed to a fixed target length (5\,s) and sample rate (32\,kHz). We distinguished the audio files into the coarse target classes and silence based on their fine-granular labels and tagged them accordingly to leverage the data for our purpose. The scripts for mixing the data can be found in our repository.

The mixing process for our three main classes and their combinations is as follows: 
For mixtures with a single active class (exclusive cases), we draw up to four files from that class (with two or three being most likely), apply a random per-file gain in the range $-30$ to $0$\,dB, and add them sequentially at signal-to-noise ratios (SNRs) randomly sampled from $-5$ to $+5$\,dB, normalising after each addition. For mixtures with two active classes, we sample up to three files per active class (with one per class being most likely) and combine them in the same way. For mixtures with all three classes active, we sample up to two files per class, again favouring one per class. 
Additionally, with $50\%$ probability we apply noise (white Gaussian, white uniform, or pink) to the curated audio at an SNR drawn from $-5$ to $+15$\,dB to further diversify backgrounds.
We also synthesise silent audio files for the \textbf{silence} class by creating a zero-valued waveform of the target length and sample rate and then injecting randomly chosen noise (white Gaussian, white uniform, or pink) with an initial gain between $-5$ and $+1$\,dB. Finally, a gain stage attenuation between $-40$ and $-5$ dB is applied to obtain a range of different loudness for the silence audio clips.

\subsection{Deep learning architectures}
\label{ssec:model-training}

Our first step was to benchmark a selection of popular models on {\edansa} and evaluate their transfer performance on {\besound}. To this end, we employed both \ac{CNN}-based architectures, \ie, CNN10~\citep{Kong20-PLP}, CNN14~\citep{Kong20-PLP}, ResNet-50~\citep{He16-DRL}, EfficientNet-B7~\citep{Tan19-ERM} and BirdNET~\citep{Kahl21-BAD}, as well as transformer-based architectures, \ie, AST~\citep{Gong21-AAS}, SSAST~\citep{Gong22-SSA}, PaSST~\citep{Koutini22-ETO}, AVES~\citep{Hagiwara23-AAV}, W2V2~\citep{Baevski20-W2A}, Whisper~\citep{Radford23-RSR}, CLAP-HTSAST~\citep{Wu23-LCL,Chen22-HAH}, and Qwen2-Audio~\citep{Chu24-QTR}. 

For models requiring spectrograms as input, that do not provide a dedicated feature extractor (\eg, most of the \ac{CNN}-based models), we extracted log-Mel spectrograms following the example of \citep{Kong20-PLP}, \ie, using a target sample rate of $32$\,kHz, a window size of $1024$, hop size of $320$, and $64$ mel bins. An example of the extracted spectrograms is visualised in \cref{fig:spectrograms}.

In addition, we included BirdNET as a practical baseline because of its widespread use in bioacoustic research. Unlike the other architectures, BirdNET was not fully fine-tuned using the \textit{autrainer} package. Instead, we used the publicly available BirdNET Analyzer and its default autotune workflow, following the official documentation\footnote{\url{https://birdnet-team.github.io/BirdNET-Analyzer}}. Consequently, the BirdNET results represent a practical baseline and are not directly comparable to the other architectures, which were fully fine-tuned under the same training framework.

For each model, except BirdNET, we conducted a small grid search over common training configurations. 
If available, we utilised publicly released pre-trained weights and fine-tuned them on {\edansa}. 
Typical pretraining datasets are AudioSet \citep{Gemmeke17-ASA} (\eg, CNN10, CNN14, AST, PaSST, SSAST), ImageNet \citep{Russakovsky15-ILS} (\eg, ResNet-50, EfficientNet-B7), LibriSpeech (\eg, W2V2), a combination of public audio datasets (\eg, AVES), and also a combination of large scale proprietary and public audio data (\eg, Whisper and Qwen2-Audio).

In addition to the {\edansa} test set, we evaluated all architectures on {\besound} to assess their transfer to the intended target domain.
As we are dealing with a three-class multi-label soundscape component classification task, we trained all networks with a binary cross-entropy loss, as multiple tags can be active simultaneously. A class is considered active when its confidence score exceeds $0.5$. Each model was trained for $30$ epochs, with the best model checkpoint on the validation set retained for evaluation. Since the {\besound} recordings were $60\,s$ long, we applied the models in a sliding-window manner using non-overlapping windows (\eg, six $10\,s$ windows). For each class, we aggregated the window-level predictions by taking the maximum confidence score across all windows and then applied the same $0.5$ threshold as for {\edansa}-test. Thus, performance is evaluated at the $10\,s$ clip level for {\edansa} and at the $60\,s$ recording level for {\besound}. Additionally, we evaluated whether a ``model soups'' ensembling strategy \citep{Wortsman22-MSA} further improved performance by averaging the weights of all grid search runs of each model.

Our hyperparameter options are listed in \cref{tab:hyperparameters}. 
However, not all combinations were explored exhaustively, mainly due to resource constraints.
Specifically, larger models, such as Qwen2-Audio, limited the feasible batch size compared to smaller ones. 
For example, this model only allowed a maximum batch size of $4$ on our strongest GPU (Nvidia A40). More detailed information on the parameters is provided in \ref{app:grid-search}. 
Moreover, no data augmentation techniques were applied at this stage. The purpose of this experiment was therefore not to achieve the absolute best results for every model, but rather to identify the most promising architectures to carry forward into subsequent experiments. All experiments were carried out on Nvidia A40 and RTX3090 GPUs.

\begin{table}[t]
    \centering
    \begin{tabular}{c|c}
        Parameter & Options \\
        \hline
        batch size & 4, 16, 32, 64 \\
        learning rate & $1e-3$, $1e-4$, $1e-5$ \\
        optimiser & Adam, AdamW \\
    \end{tabular}
    \caption{Hyperparameter options for the initial model training.}
    \label{tab:hyperparameters}
\end{table}

\subsection{Model refinement and analysis}
\label{ssec:model-refinement-and-analysis}
This subsection covers different approaches we analysed in order to improve model performance. For this, we first utilised the three most suitable models from \cref{ssec:model-training} in \cref{ssec:silence}, before selecting the final architecture for the last investigations in \cref{ssec:additional-data,ssec:evaluation-strategy}.

\subsubsection{The role of silence}
\label{ssec:silence}
In some ecoacoustic studies \citep{Quinn22-SCW,Grinfeder22-SDO,Coban22-ETE,Zhang23-COC}, the commonly used classification of one or more of the soundscape classes anthropophony, biophony, and geophony has been extended by introducing additional categories that account for silent recordings or background noise. However, the approaches and implementations differ across studies, both in terms of the number of additional classes and the definition of these classes.
For instance, \citet{Coban22-ETE} simply use a silence tag for their annotations. In this case, however, silence is sometimes intermingled with geophonic events (\eg, rain or wind), which blurs the conceptual boundary between these categories and therefore does not strictly distinguish between them. 
In other studies, quiet recordings or silence is either not considered \citep{Challeat24-ADO,Fairbrass19-CDL,Ferreira25-TMI} or interpreted as part of geophony \citep{Wang25-RDD}.
This lack of a clear distinction complicates interpretation and can introduce inconsistencies across datasets and models.

We assume that if none of our three target classes is recognised, the recording can be reasonably considered as \textit{silence}. This definition tries to avoid overlap with geophony and ensures that silence is treated as an absence of acoustic events. 

Therefore, this line of experiments investigated how including silence as a fourth target class during model training affects the model performance on our three primary classes. While the silence class was included during training and validation, it was omitted from the test evaluation, as our primary objective was to assess the performance on the three target classes.

We conducted the experiment on the top-3 models from \cref{ssec:model-training}, fine-tuning them on {\edansa} and evaluating them on the test set from {\edansa} as well as our main target domain set {\besound}. The training pipeline is the same as before, but with adjusted hyperparameter options for the grid search, as listed in \cref{tab:hyperparameters-silence} and details in \ref{app:grid-search}. Furthermore, we stuck to Adam as optimiser but investigated more learning rates and also applied data augmentation as we reduced our models to the top-3 and focus our resources on those. In this context, we had three different setups: 1) no data augmentation as before, 2) SpecAugment \citep{Park19-SAS}, or 3) a custom augmentation pipeline. The custom augmentation pipeline applies either Gaussian noise or SpecAugment to the training samples, chosen with a probability of $30$\,\% and $70$\,\%, respectively. The chosen augmentation is then applied with $80$\,\% probability. The design emphasizes spectrogram masking for robustness while still incorporating occasional noise and clean examples. Moreover, we also conducted balanced sampling, \ie, for each training batch of audio samples we try to sample the classes that have less samples more often than the categories with many samples by assigning the classes respective weights.

\begin{table}[]
    \centering
    \begin{tabular}{l|c}
        Parameter & Options \\
        \hline
        batch size & $16$, $32$ \\
        learning rate & $1e\!-\!3$, $5e\!-\!4$, $1e\!-\!4$, $1e\!-\!5$ \\
        optimiser & Adam \\
        augmentation & None, SpecAugment, CustomPipeline \\
    \end{tabular}
    \caption{Possible hyperparameter options for the model training with silence.}
    \label{tab:hyperparameters-silence}
\end{table}

\subsubsection{Impact of additional training data}
\label{ssec:additional-data}

A common paradigm in deep learning, especially in times of foundation models and \acp{LLM}, is that more data leads to better performance \citep{Kaplan20-SLF,Hoffmann22-TCL}. Thus, we hope to bridge the \textit{domain gap} by adding more diverse and also more similar data (\wrt our target domain). In this experiment, we investigated whether more data improves model generalisation in our eco-acoustic setting, and in particular which of our data sources or their combinations lead to the best model performance. In this context, we drew on the additional data sources introduced in \cref{ssec:data}: {\akwamo}, {\beambient}, {\htsforest}, and {\synthdata}. Now deploying the best performing model from \cref{ssec:silence}, we retrained the model with its corresponding configuration and two different learning rates ($1e\!-\!4$, $1e\!-\!5$) to allow for some adaptability to the data sources. Each training run combined the base dataset {\edansa} with one or more of the four supplementary datasets. In particular, the various settings can be summarised as follows: 
\begin{itemize}
    \item \textbf{Single-dataset addition:} each dataset is added individually to {\edansa}, without any other dataset.
    \item \textbf{Regionally similar datasets:} both {\beambient} and {\htsforest}, which were recorded in similar \ac{BE} environments, are added together.
    \item \textbf{All \ac{PAM} datasets:} {\beambient}, {\htsforest}, and {\akwamo} are jointly included.
    \item \textbf{All datasets:} All four datasets, including {\synthdata}, are added to {\edansa}.
\end{itemize}

We considered performance on both {\edansa} and {\besound} when comparing training-data compositions, with the aim of improving robustness across \ac{PAM} domains while maintaining strong performance under the recording conditions relevant to the BE.

\subsubsection{Evaluation strategy and target-domain calibration}
\label{ssec:evaluation-strategy}
As demonstrated in several bio- or ecoacoustic studies \citep{Scanferla25-DST,Wood24-GFA,Arend25-SEO,Tseng25-SBC,Funosas26-AGA}, tailoring thresholding and evaluation strategies to the specific target classes and domain data can substantially improve model performance. This can, for instance, be achieved by applying class-specific prediction thresholds instead of using a single global threshold across all target categories \citep{Scanferla25-DST,Tseng25-SBC}, or by employing a count-based thresholding approach, \ie, requiring a certain number of prediction windows to exceed a probability threshold \citep{Arend25-SEO}, which might be beneficial for long audio recordings.

Accordingly, the experiments in this subsection aim to enhance performance on our {\besound} dataset by exploring 1) duration-based annotation adaptation, 2) class-specific thresholds, 3) the combination of both, and 4) count-based thresholding.

These optimisations constitute post-hoc target-domain calibration and alternative task operationalisations on {\besound} and are applied without additional model training. As {\besound} annotations are used, \eg, to select class-specific decision thresholds and duration-based criteria, the resulting performance should be interpreted as a domain-calibrated upper-bound estimate rather than an unbiased estimate of generalisation to unseen recordings. The subsequent error analysis and ecological case study in \cref{ssec:results-error-analysis,ssec:results-coarsenet-vs-indices} therefore used the calibrated model.

\textbf{\Ac{PDA}:} This approach tries to mitigate the influence of falsely annotated or less representative labels. To this end, we exclude labels with very short temporal duration. While biophony can indeed exhibit very short event times, geophony is typically longer (\eg, weather events such as wind or rain typically occur across longer time spans like $30\,s$ to hours) as well as anthropophony (as the focus lies on technophony)~\citep{Farina21-POT}. 
As we are primarily interested in wind and rain, the relevant geophonic events in our study are expected to be predominantly of ephemeral or continuous nature~\citep{Farina21-POT}. Similarly, the anthropophony of interest consists mainly of technophonic events, such as highway traffic or aircraft overflights, which are typically ephemeral or continuous rather than abrupt~\citep{Farina21-POT}. Consequently, very short annotations may be less representative of the target class or may reflect annotation uncertainty and should therefore be treated with caution.

For this purpose, we leverage the strongly annotated labels of {\besound}. Since every recording is $60\,s$ long, we look at the duration of each annotation of every target class. We expect each annotation to be at least $t$ seconds long, where $t$ is supposed to be $p$ percent of the full recording length $T$ (\ie, for us $60\,s$) with $p \in \{.05, .10, .25\}$, \ie, for $p = .05$ we obtain a minimal required annotation duration of $3\,s$. For now, we stick with the global threshold of $0.5$ and choose the $p$ achieving the best F1-score for each target class individually, except biophony which simply uses the normal annotations. In this context, we will also report the macro F1-score, \ie, the unweighted average across all target classes.

\textbf{\Ac{CST}:} Here, we determine the best individual threshold for each of the three main categories, \ie, the minimum value that needs to be exceeded for a class to be considered active. The final threshold for each class is the one yielding the highest F1-score on {\besound}. We do that in order to obtain an upper bound of performance. 

\textbf{Combination of \ac{PDA} and \ac{CST}:} This method combines the \ac{PDA} and \ac{CST} approach by using the adapted annotations for anthropophony and geophony and finding the best class-specific threshold for each class. For this, we use \ac{PR} as well as \ac{ROC} curves in order to visualise the behaviour per class. We choose the best threshold per class based on the best F1-score (\wrt the \ac{PR} curve) and Youden's index (\wrt to the \ac{ROC} curves).

\textbf{\Ac{CBT}:} In this approach, we investigate how many prediction windows must exceed the class-specific threshold for a category to be considered active. This builds on top of the \ac{PDA} + \ac{CST} method previously described. To enable a more fine-grained analysis, the inference step size is reduced from $10\,s$ (\ie, $6$ prediction windows in total) to $1\,s$, leading to $51$ prediction windows. For each class, we then search for the optimal count $c$, which achieves the maximum F1-score on {\besound}. Specifically, we consider different percentages $p \in \{.05, .10, .15, .20, .25\}$ of the total number of prediction windows ($w = 51$). The corresponding count is computed as $c = \lfloor p \cdot w\rfloor$. For instance, with $p = .05$, we obtain $c = 2$. In the end, a class is predicted as active only if at least $c$ prediction windows exceed the corresponding CST.

\subsection{Ecological case study}
\label{ssec:methodology-ecological-case-study}

In this case study, we investigate the effectiveness of {\coarsenet} as a pre-processing step in a standard ecoacoustic analysis workflow.
It has been shown that standard ecoacoustic indices can correlate with ecological indicators, such as the $\alpha$-diversity of avian species richness, depending on the context and with careful interpretation~\citep{Towsey14-TUO,Droege21-LTA,Eldridge18-SOE,Bradfer20-RAO,Shaw24-FSH}. As this assumption has primarily been demonstrated in temperate regions \citep{Eldridge18-SOE,Shaw24-FSH}, it is reasonable to expect that it also applies to the \ac{BE} regions, which are also temperate.
To that end, we manually annotated a subset of our {\besound} data ($852$ recordings) for bird species and computed the $\alpha$-diversity, which we defined here as the number of species in each recording.
Subsequently, we computed three standard ecoacoustic indices, namely, ADI \citep{Villanueva11-APO}, \ac{ACI} \citep{Pieretti11-ANM}, and \ac{NDSI} \citep{Kasten12-TRE}. \citet{Bradfer19-GFT} describe those indices and their purpose in more detail. 
We then correlated each of them with $\alpha$-diversity using Pearson's correlation coefficient once for all data ($x \in A \cup B \cup G$) and then for filtered versions thereof, where we first considered data containing only biophonic sounds ($x \in B$) and then data which also include anthropophony ($x \in A \cup B$) or geophony ($x \in B \cup G$).
This corresponds to the use of {\coarsenet} to limit the analysis on only ``clean'' audio data, \ie, only containing biophonic sounds, or one with some contamination (but only from one source).
To get an upper bound on performance, we also filtered using the ground truth human annotations.

\section{Results}
\label{sec:results}

\subsection{Deep learning architectures}
\label{ssec:results-model-training}

The results for the initial benchmarking of a selection of popular deep learning architectures on {\edansa}-test and {\besound} are shown in \cref{tab:edansa-model-results} and \cref{tab:besound-initial-model}. On {\edansa}-test, pre-trained \ac{CNN}-based architectures performed best, with CNN10 achieving the highest macro F1-score. However, AST, a transformer-based model, was highly competitive and tied for second place with CNN14.
In contrast, on the {\besound} dataset the strongest performing models were transformer-based, particularly large foundation model encoders, with Qwen2-Audio achieving the best results. 

Across all models, we observe a performance drop for all our target categories compared to {\edansa}. This especially applies to anthropophony, followed by geophony, while biophony remains relatively stable. This suggests a domain gap between the two datasets, such that patterns learnt on {\edansa} do not well transfer to {\besound}. 
To address this gap and improve cross-domain generalisation, while keeping resource requirements feasible, we selected three models for further experiments: CNN10, AST, and CLAP-HTSAST. 
We chose these models based on their performance on both the EDANSA-test and {\besound} datasets.
CNN10 and AST combine high performance with fast inference on the {\edansa}-test, giving us both a \ac{CNN}- and a transformer-based option. CLAP-HTSAST was chosen for its strong performance on {\besound} and substantially faster inference compared to Qwen2-Audio.

Overall, the architecture comparison revealed substantial performance differences across \ac{PAM} domains, highlighting the importance of considering cross-domain transfer during model selection.

\begin{table}[t]
\centering
\caption{Clip-level multi-label classification performance on the {\edansa} test set using F1-score as evaluation metric. The best performance is marked \textbf{bold} while the second best is \underline{underlined} and the third best is in \textit{italic} font. The macro F1 score is reported together with the corresponding \ac{CI} of 95\%. Furthermore, the \ac{MIT} for a 60\,s long audio file, utilising a sliding window of size 10\,s and a step size of 10\,s, is reported.
}
\label{tab:edansa-model-results}
\resizebox{\textwidth}{!}{%
\begin{tabular}{l|ccc|c|c}
\textbf{Model} & \textbf{Anth} & \textbf{Bio} & \textbf{Geo} $\uparrow$ & \textbf{Macro F1} $\uparrow$ & \textbf{MIT (s)} $\downarrow$ \\
\hline
CNN 10 ($lr=.001$, $bs=32$, Adam) & .954 & .956 & .865 & \textbf{.925}, CI $[.918, .932]$ & .071 \\
CNN 14 ($lr=.0001$, $bs=64$, AdamW) & .942 & .956 & .865 & \underline{.921}, CI [.913, .928] & .078 \\
ResNet-50 ($lr=.001$, $bs=32$, Adam) & .951 & .948 & .860 & \textit{.920}, CI [.912, .927] & .100 \\
EfficientNet-B7 ($lr=.001$, $bs=64$, AdamW) & .948 & .947 & .847 & .914, CI [.906, .921] & .230 \\
BirdNET (using autotune\footnotemark) & .871 & .911 & .747 & .843, CI [.833, .853] & -- \\
AST ($lr=.0001$, $bs=16$, Adam) & .966 & .949 & .849 & \underline{.921}, CI [.914, .927] & .117 \\
SSAST ($lr=.00001$, $bs=16$, Adam) & .962 & .940 & .821 & .908, CI [.901, .916] & .061 \\
PaSST ($lr=.0001$, $bs=16$, AdamW, FullFT) & .949 & .933 & .846 & .909, CI [.901, .917] & .131 \\
AVES ($lr=.0001$, $bs=16$, Adam, FullFT) & .932 & .931 & .832 & .898, CI [.891, .907] & .065 \\
W2V2 ($lr=.00001$, $bs=16$, Adam) & .902 & .896 & .813 & .870, CI [.861, .879] & .068 \\
Whisper ($lr=.00001$, $bs=16$, AdamW) & .937 & .950 & .833 & .907, CI [.899, .914] & .207 \\
CLAP ($lr=.00001$, $bs=16$, Adam, FullFT) & .940 & .954 & .795 & .897, CI [.888, .905] & .195 \\
Qwen2-Audio ($lr=.00001$, $bs=4$, AdamW) & .945 & .947 & .832 & .908, CI [.900, .916] & .756 \\
\end{tabular}
}
\end{table}

\begin{table}[t]
\centering
\label{tab:besound-initial-model}
\caption{
Recording-level multi-label classification results on {\besound} using a window size of $10$\,s and a step size of $10$\,s. Once a category had a prediction of $>.5$ in at least one prediction window, the respective class was considered active for the current recording. The macro F1 score is reported together with the corresponding \ac{CI} of 95\%. The best performing model is marked \textbf{bold}, the second best \underline{underlined}, and the third best \textit{italic}.}
\resizebox{.6\textwidth}{!}{%
\begin{tabular}{l|ccc|c}
\hline
\textbf{Model} & \textbf{Anth} & \textbf{Bio} & \textbf{Geo} & \textbf{Macro} $\uparrow$\\
\hline
CNN 10          & .199 & .893 & .623 & .571, CI [.561, .583] \\
CNN 14          & .227 & .920 & .621 & .589, CI [.579, .601] \\
ResNet-50       & .223 & .862 & .616 & .567, CI [.557, .579] \\
EfficientNet-B7 & .483 & .866 & .636 & .662, CI [.652, .671] \\
BirdNET         & .327 & .803 & .599 & .576, CI [.565, .586] \\
AST             & .421 & .872 & .659 & .651, CI [.640, .661] \\
SSAST           & .190 & .880 & .634 & .568, CI [.558, .580] \\
PaSST           & .251 & .890 & .662 & .601, CI [.590, .612] \\
AVES            & .546 & .869 & .611 & .675, CI [.665, .685] \\
W2V2            & .551 & .829 & .652 & .677, CI [.668, .686] \\
Whisper         & .456 & .900 & .681 & \textit{.679}, CI [.670, .688] \\
CLAP            & .512 & .910 & .631 & \underline{.684}, CI [.674, .695] \\
Qwen2-Audio     & .557 & .898 & .669 & \textbf{.708}, CI [.699, .717] \\
\hline
\end{tabular}
}
\end{table}

\subsection{The role of silence}
\label{ssec:results-silence}

The model performance of our three chosen models based on \cref{ssec:results-model-training}, now also including a silence category during training, are listed in \cref{tab:top3-edansa-results} on the {\edansa}-test and \cref{tab:besound-silence-results} on {\besound}. On {\edansa}, the models perform better when excluding the silence class from training, which is reasonable considering that some samples of silence are annotated together with geophonic events.
In contrast, the model performance on the target categories on {\besound} increased for all three models when including silence as an additional class during training. Since the {\besound} data has more similarities with the application area of our model (\ie, the \ac{BE} regions), we decide to include the silence category for the succeeding experiments. However, we will not use the model prediction of silence or evaluate it, but simply annotate silence to audio samples where none of the target categories (anthropophony, biophony, geophony) is predicted. 

\begin{table}[t]
\centering
\caption{Clip-level multi-label classification performance on the {\edansa} test set, when including the silence class during model training.}
\label{tab:top3-edansa-results}
\resizebox{\textwidth}{!}{%
\begin{tabular}{l|l|ccc|c}
\textbf{Model} & \textbf{Training Setup} & \textbf{Anth} & \textbf{Bio} & \textbf{Geo} $\uparrow$ & \textbf{Macro F1} $\uparrow$ \\
\hline
CNN 10 ($bs=16$, $lr=.001$, Adam, aug=None, pre-trained) & Without Silence & .956 & .951 & .861 & \textbf{.923}, CI [.915, .930]\\
CNN 10 ($bs=16$, $lr=.001$, Adam, aug=SpecAugment) & With Silence    & .960 & .953 & .843 & .918, CI [.911, .925]\\ 
AST ($bs=16$, $lr=.0001$, Adam, aug=None)    & Without Silence & .956 & .943 & .857 & .919, CI [.911, .926]\\
AST ($bs=16$, $lr=1e-05$, Adam, aug=CustomPipeline)    & With Silence    & .919 & .957 & .852 & .909, CI [.901, .918]\\ 
CLAP ($bs=16$, $lr=.0001$, Adam, aug=SpecAugment)   & Without Silence & .935 & .953 & .882 & .923, CI [.916, .931]\\ 
CLAP ($bs=16$, $lr=1e-05$, Adam, aug=SpecAugment)   & With Silence    & .951 & .952 & .819 & .907, CI [.899, .915]\\ 
\end{tabular}
}
\end{table}

\footnotetext{We stuck to the official documentation from \url{https://birdnet-team.github.io/BirdNET-Analyzer} and used the default autotune settings.}

\begin{table}[t]
\centering
\caption{Recording-level multi-label classification performance on the {\besound} set, when including the silence category during model training. A window size of $10$\,s and a step size of $10$\,s were applied. Once a category had a prediction of $>.5$, in at least one of the prediction windows, the respective class was considered active for the current recording. The macro F1 score is reported together with the corresponding \ac{CI} of 95\%.}
\label{tab:besound-silence-results}
\resizebox{\textwidth}{!}{%
\begin{tabular}{l|l|ccc|c}
\textbf{Model} & \textbf{Training Setup} & \textbf{Anth} & \textbf{Bio} & \textbf{Geo} $\uparrow$ & \textbf{Macro F1} $\uparrow$ \\
\hline
CNN 10 & Without Silence & .250 & .880 & .685 & .605, CI [.594, .616]\\
CNN 10 & With Silence    & .316 & .911 & .690 & .639, CI [.629, .649] \\
AST    & Without Silence & .385 & .868 & .539 & .597, CI [.585, .609] \\
AST   & With Silence    & .291 & .908 & .657 & .619, CI [.608, .630] \\
CLAP   & Without Silence & .308 & .904 & .676 & .629, CI [.619, .640] \\
CLAP   & With Silence    & .479 & .913 & .658 & .683, CI [.673, .693] \\
\end{tabular}
}
\end{table}

\subsection{Impact of additional training data}
\label{ssec:results-adding-data}

The results in \cref{tab:data-variations-edansa} suggest that the upper limit of {\edansa}-test has been reached and no additional performance boost is achieved by adding more data. In contrast, the performance on {\besound} in \cref{tab:data-variations-besound} could be improved with some data combinations. However, for the addition of only one additional dataset to {\edansa} only {\beambient} could achieve a better performance than without. For all other scenarios, where at least two datasets were added, the model performance increased as well. The best performance boost could be achieved by adding all the \ac{PAM} datasets. 
The second best place is taken by adding the two datasets which were also collected on the \ac{BE} regions, just as {\besound}. 
In contrast, adding the mixed data ({\synthdata}) achieved the worst performance. 

Ultimately, these results indicate that training-data composition affects model performance, with domain-relevant \ac{PAM} data providing the greatest benefit.

\begin{table}[t]
    \centering
    \caption{Datasets added to {\edansa} during model training. Clip-level multi-label classification results on {\edansa} test. The macro F1 score is reported together with the corresponding \ac{CI} of 95\%. The best result is marked \textbf{bold}.}
    \resizebox{\textwidth}{!}{%
    \begin{tabular}{l|ccc|c}
         Added Dataset(s) & Anth & Bio & Geo & Macro $\uparrow$\\
         \hline
         Baseline & .951 & .952 & .819 & .907, CI [.899, .915] \\
         \hline
         \htsforest & .946 & .955 & .866 & \textbf{.922}, CI [.915, .930] \\
         \beambient & .944 & .952 & .815 & .904, CI [.895, .911] \\ 
         \akwamo & .850 & .954 & .804 & .869, CI [.859, .880] \\
         \synthdata & .929 & .955 & .773 & .886, CI [.876, .895] \\ 
         \hline
         BE-data (\htsforest, \beambient) & .916 & .957 & .817 & .897, CI [.888, .905] \\ 
         PAM-data (\htsforest, \beambient, \akwamo) & .930 & .953 & .865 & .916, CI [.908, .923] \\ 
         ALL-data (all four above) & .903 & .958 & .809 & .890, CI [.880, .900] \\
    \end{tabular}
    }
    \label{tab:data-variations-edansa}
\end{table}

\begin{table}[t]
    \centering
    \caption{Datasets added to {\edansa} during model training. Recording-level multi-label classification results on {\besound} with 
    a window size of $10$\,s and a step size of $10$\,s. The macro F1 score is reported together with the corresponding \ac{CI} of 95\%. The best result is marked \textbf{bold}.}
    \resizebox{\textwidth}{!}{%
    \begin{tabular}{l|ccc|c}
         Added Dataset(s) & Anth & Bio & Geo & Macro $\uparrow$\\
         \hline
         Baseline & .479 & .913 & .658 & .683, CI [.673, .693] \\
         \hline
         \htsforest & .516 & .886 & .672 & .691, CI [.683, .700] \\
         \beambient & .619 & .907 & .694 & .740, CI [.731, .749] \\
         \akwamo & .382 & .905 & .692 & .660, CI [.650, .670] \\
         \synthdata & .491 & .864 & .610 & .655, CI [.644, .664] \\
         \hline
         BE-data (\htsforest, \beambient) & .634 & .906 & .720 & .753, CI [.744, .762] \\
         PAM-data (\htsforest, \beambient, \akwamo) & .649 & .909 & .717 & \textbf{.758}, CI [.750, .767] \\
         ALL-data (all four above) & .635 & .914 & .691 & .747, CI [.738, .756] \\
    \end{tabular}
    }
    \label{tab:data-variations-besound}
\end{table}

\subsection{Evaluation strategy and target-domain calibration}
\label{ssec:results-evaluation-strategy}

The results of the different approaches are summarised in \cref{tab:thresholding}. As described in \cref{ssec:evaluation-strategy}, these results are optimised on {\besound} itself and therefore represent domain-calibrated rather than independent generalisation performance.
For \ac{PDA}, the best $p$ for anthropophony was $p>0$, \ie, using the baseline, while for geophony, it was $p=5$. However, when applying \ac{PDA} with a global confidence threshold, an improvement over the baseline was observed only for geophony.
On the other hand, the performance improved across all target classes when deploying \ac{CST}, \ie, using a class-specific confidence threshold, although the improvement for geophony was only marginal ($+.002$). The corresponding confidence thresholds were $.722$ for anthropophony, $.920$ for biophony, and $.571$ for geophony.  

\begin{figure}[t]
    \centering
    \includegraphics[width=\linewidth]{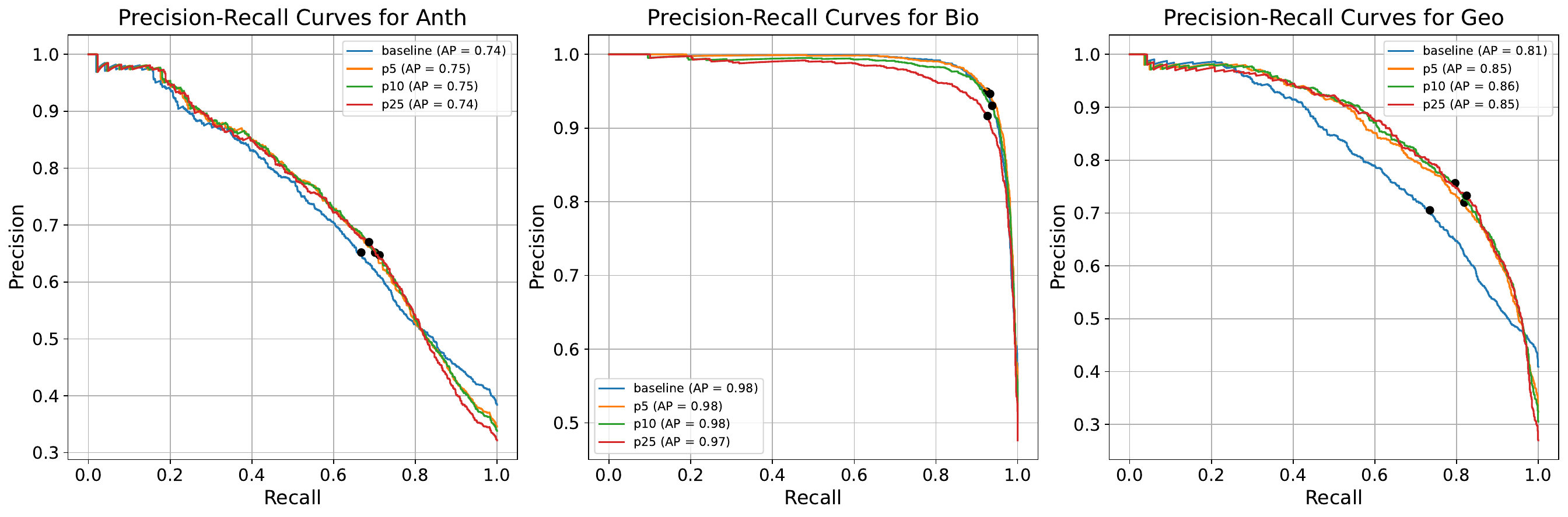}
    \caption{The \ac{PR} curves for the three classes Anthropophony (Anth), Biophony (Bio), and Geophony (Geo). Every class has four curves, representing the thresholding percentages described in \cref{ssec:evaluation-strategy}, \ie, 5\,\%, 10\,\%, 25\,\% of the full recording length, and the baseline.}
    \label{fig:pr-curve-th-percentages-max}
\end{figure}

\begin{figure}[t]
    \centering
    \includegraphics[width=\linewidth]{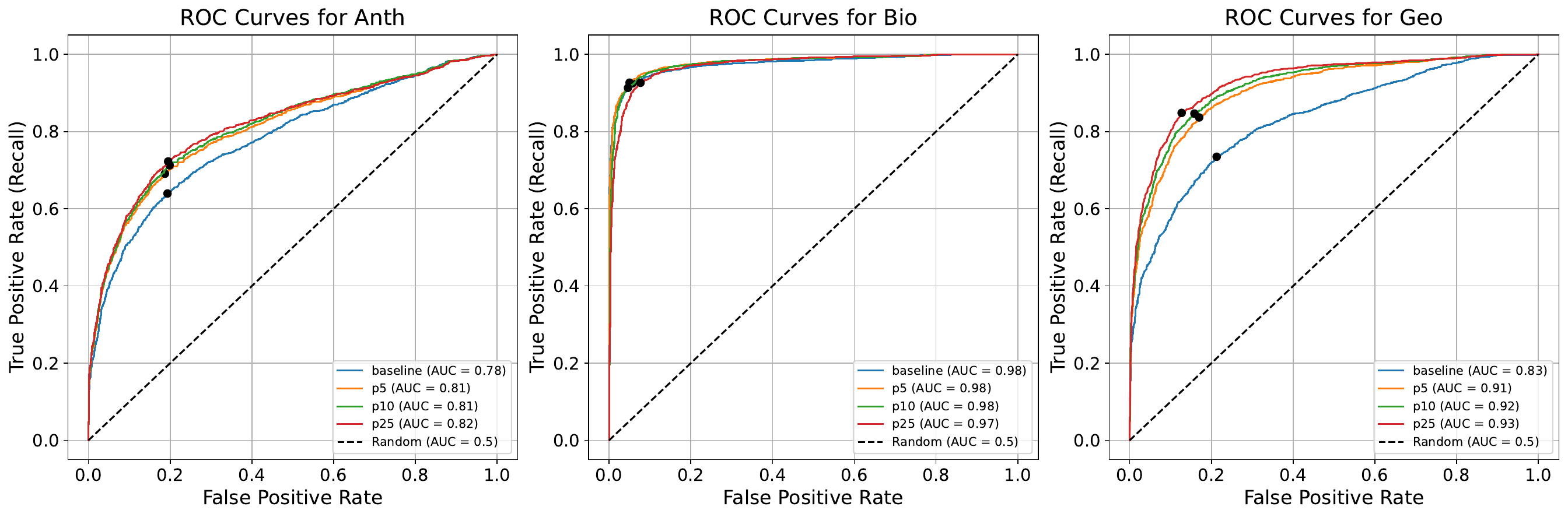}
    \caption{The \ac{ROC} curves for the three classes Anthropophony (Anth), Biophony (Bio), and Geophony (Geo). Every class has four curves, representing the thresholding percentages described in \cref{ssec:evaluation-strategy}, \ie, 5\,\%, 10\,\%, 25\,\% of the full recording length, and the baseline.}
    \label{fig:roc-curve-th-percentages-max}
\end{figure}

When combining the two approaches, the performance of every category could be increased noticeably, with the biggest improvement for geophony, achieving the second best macro F1-score out of the five investigated approaches. 
The corresponding class-specific \ac{PR} and \ac{ROC} curves are provided in \cref{fig:pr-curve-th-percentages-max,fig:roc-curve-th-percentages-max}.
Here, the best results were achieved with applying $p=25$ (\ie, $15\,s$) as \ac{PDA} for anthropophony and geophony and class-specific thresholds of $.835$, $.920$, and $.927$ for anthropophony, biophony, and geophony, respectively. 
We additionally assessed the post-hoc calibration of the three aforementioned variants using three-fold cross-validation on {\besound} \cref{tab:threshold_variants_cv}.

The best macro F1-score was achieved by applying \ac{PDA} + \ac{CST} + count-based, reaching a macro F1-score of $.799$. For this, we utilised counts of $c=2$, $c=5$, and $c=10$ for anthropophony, biophony, and geophony, respectively.
However, this result is only marginally better ($+.002$) than the \ac{PDA} + \ac{CST} method, yielding almost the same performance. Considering the inference times reported in \cref{tab:edansa-model-results}, we recommend applying the \ac{PDA} + \ac{CST} approach, as it is substantially faster with only $6$ prediction windows needed instead of $51$. As a consequence, we adopt the \ac{PDA} + \ac{CST} variant for all remaining analyses regarding the error analysis and the ecological case study. 

In summary, the post-hoc experiments show that target-domain performance can be further improved through class-specific threshold calibration and, where appropriate, duration-based annotation constraints. An overview of the general {\coarsenet} model configuration and the recommended post-hoc configurations for {\besound} is provided in \cref{tab:final-configuration}.

\begin{table}[t]
    \centering
    \caption{The recording-level multi-label classification results for the different thresholding versions from \cref{ssec:evaluation-strategy}. The upper part of the table shows the results for the maximum confidence score (MCS) versions while the lower part presents the count-based results, both described in \cref{ssec:evaluation-strategy}. The macro F1 score is reported together with the corresponding \ac{CI} of 95\%.}
    \resizebox{\textwidth}{!}{%
    \begin{tabular}{l|ccc|c}
         Threshold variant & Anth & Bio & Geo & Macro $\uparrow$\\
         \hline
         Baseline & .649 & .909 & .717 & .758, CI [.750, .767] \\ 
         \ac{PDA} & .649 & .909 & .740 & .766, CI [.757, .774] \\
         \ac{CST} & .659 & .937 & .719 & .772, CI [.763, .780] \\
         \ac{PDA} + \ac{CST} & .678 & .937 & .776 & \underline{.797}, CI [.787, .805] \\
         \hline
         \ac{PDA} + \ac{CST} + count-based & .678 & .936 & .782 & \textbf{.799}, CI [.789, .808] \\
    \end{tabular}
    }
    \label{tab:thresholding}
\end{table}

\subsection{Error analysis}
\label{ssec:results-error-analysis}
 
\begin{figure}[t]
    \centering
    \includegraphics[width=.33\textwidth]{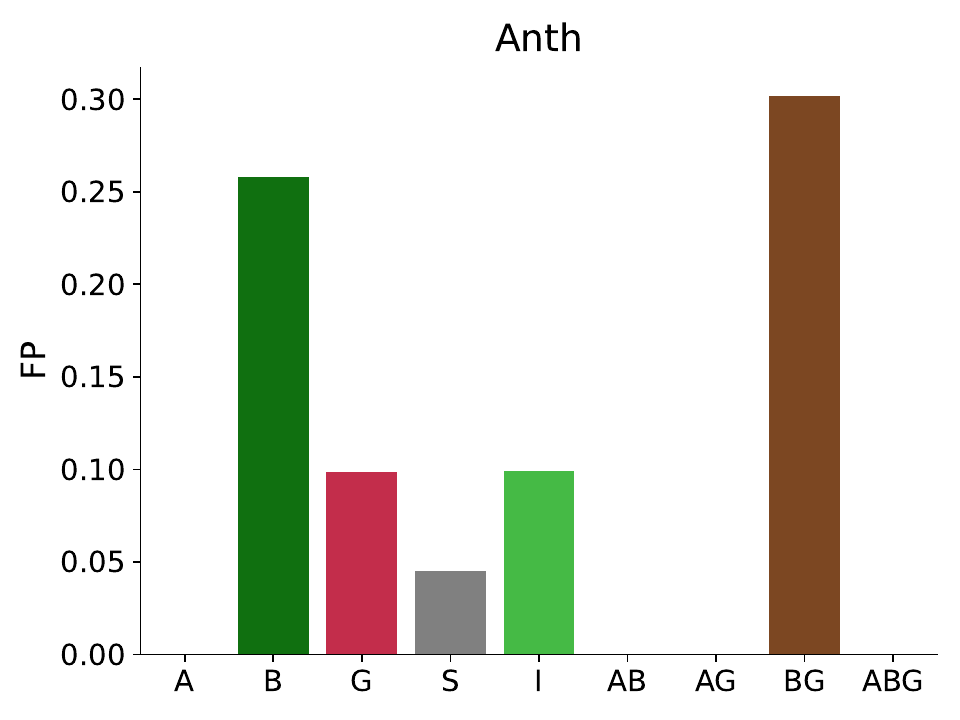}~%
    \includegraphics[width=.33\textwidth]{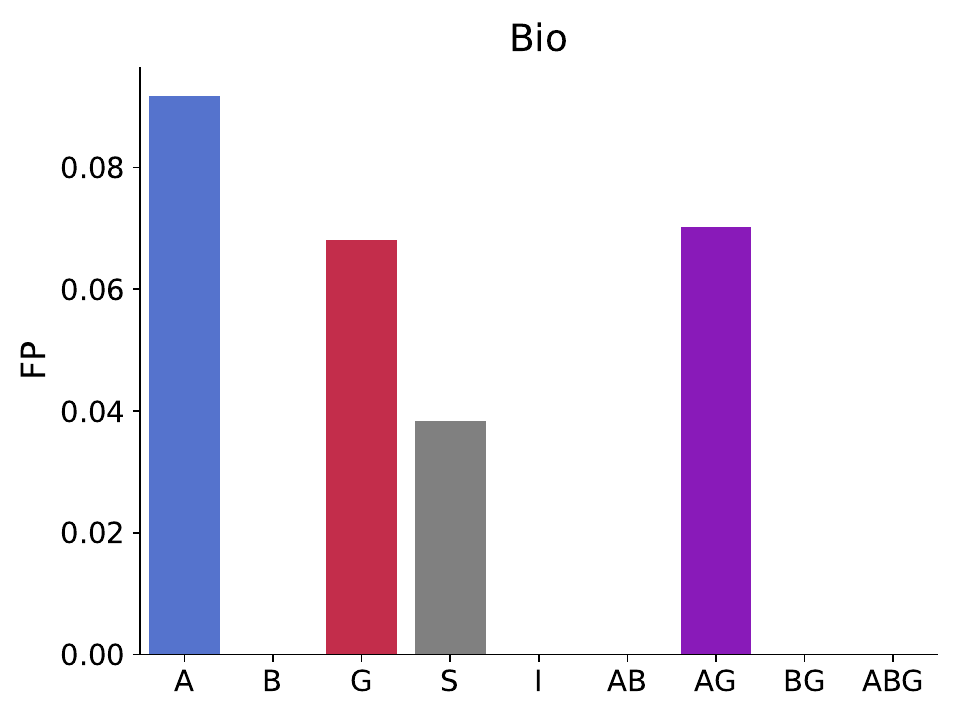}~%
    \includegraphics[width=.33\textwidth]{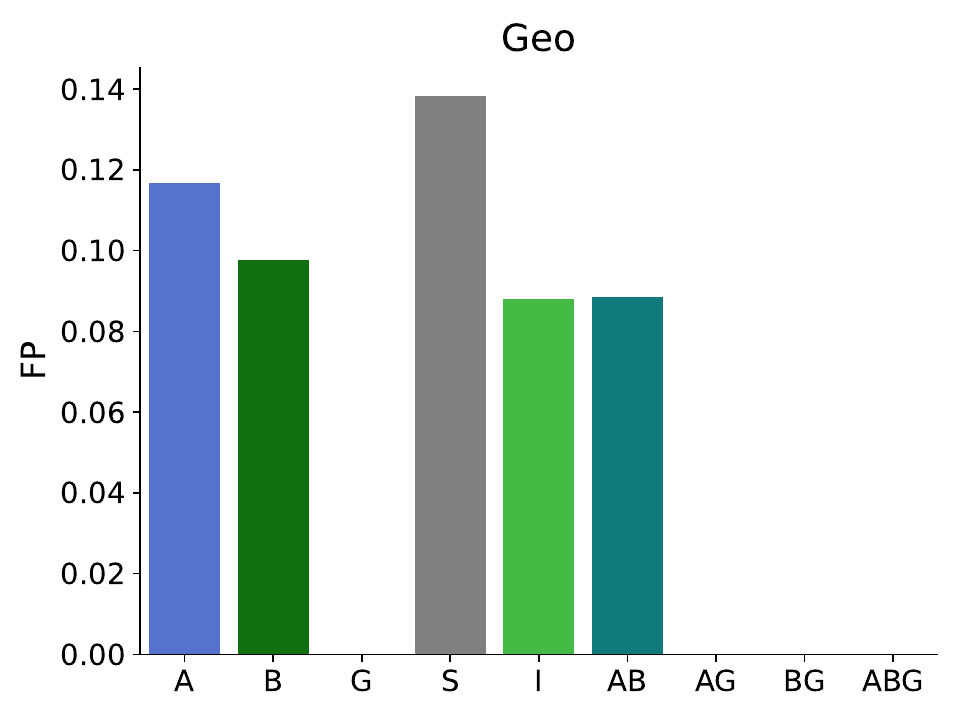}

    \includegraphics[width=.33\textwidth]{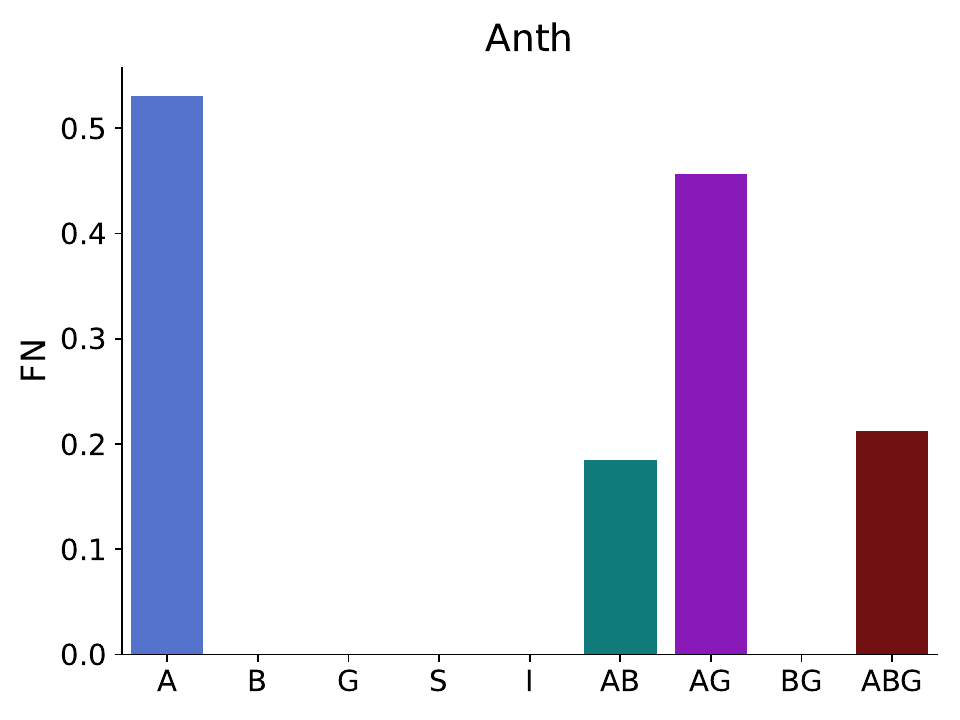}~%
    \includegraphics[width=.33\textwidth]{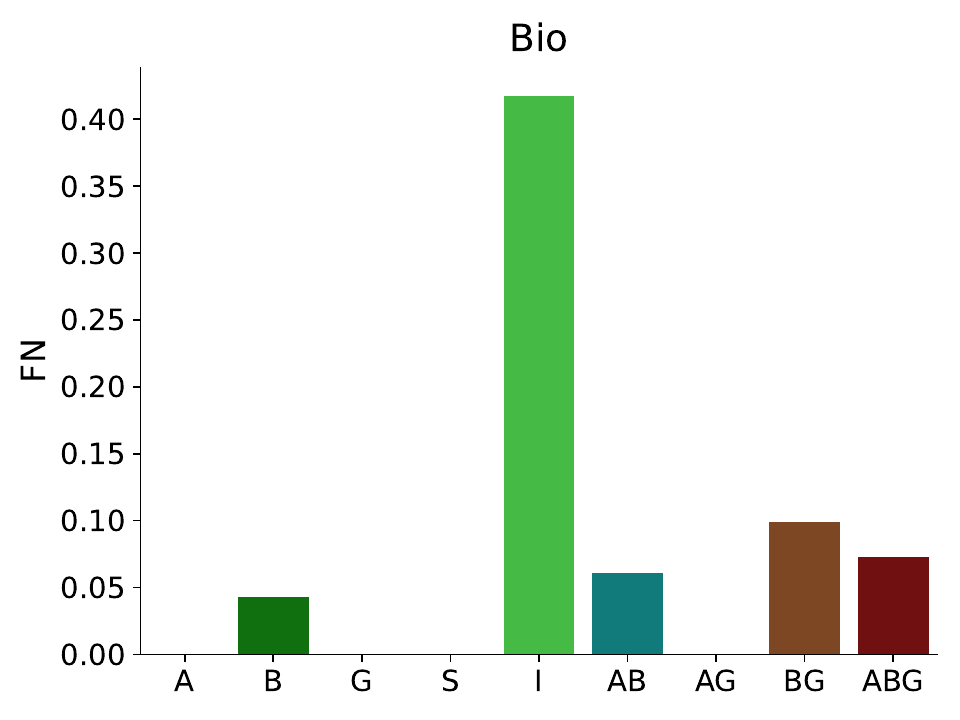}~%
    \includegraphics[width=.33\textwidth]{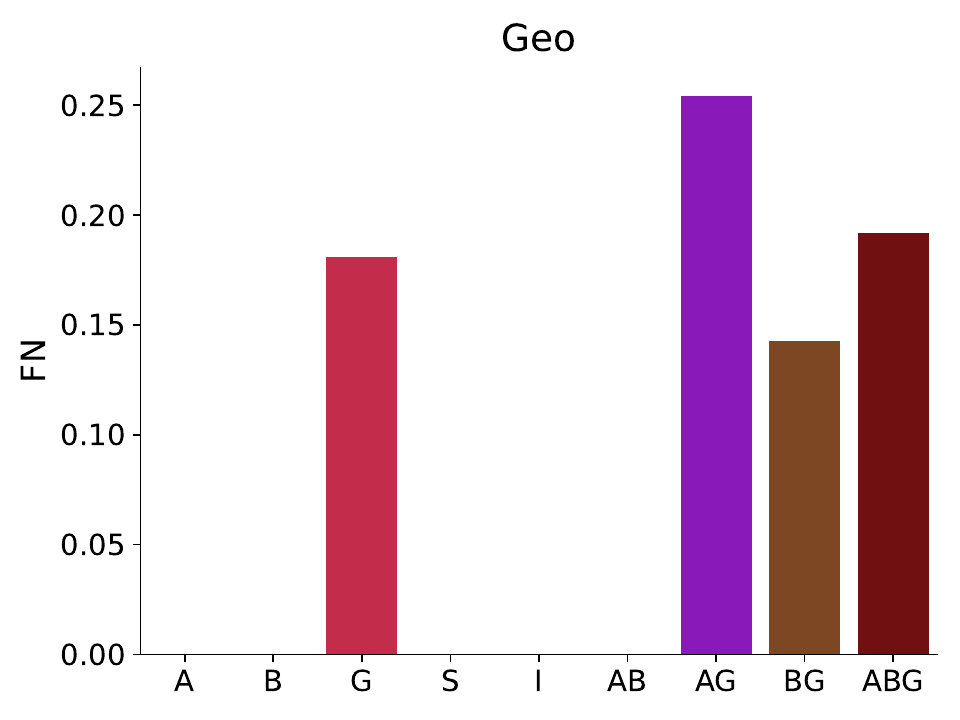}
    \caption{
    False positives (FPs; top row) and false negatives (FNs; bottom row) for the predictions of ABG stratified according to the presence of other labels. 
    }
    \label{fig:label-interaction}
\end{figure}

To gain further insights into the model's errors and confusions,
the false positives (FPs, top row) and false negatives (FNs, bottom row) are visualised in \cref{fig:label-interaction}, stratified by which other labels are annotated as active in the same segments.  
For anthropophony, most FPs occur when biophony (B) is present and especially when biophony as well as geophony are both annotated (BG). 

Regarding biophony, the overall FP rate is quite low, reflecting the strong performance on this class. The FN plot further affirms its robust performance. However, the majority of biophony FNs are observed on recordings with insect sounds. 
Furthermore, we observe some FNs when biophony co-occurs with geophony (BG), which may be attributed to strong geophonic events overlaying the biophonic activity in certain recordings.

Considering geophony, FPs mostly appear in recordings labelled as silence. Especially as wind can easily be confused with background or microphone noise, this makes sense. \citet{Coban22-ETE} sometimes even tag geophony together with silence. Though, the FNs show a different pattern, as the highest FN rates occur when geophony co-occurs with one or more of the other classes (AG, BG, ABG). The FN is particularly high when anthropophony is also active in a recording, suggesting some kind of masking effect where anthropophonic sounds dominate and suppress the detectability of geophonic events.

\subsection{CoarseSoundNet vs Ecoacoustic Indices}
\label{ssec:results-coarsenet-vs-indices}

\begin{figure}[t]
    \centering
    \includegraphics[width=.33\textwidth]{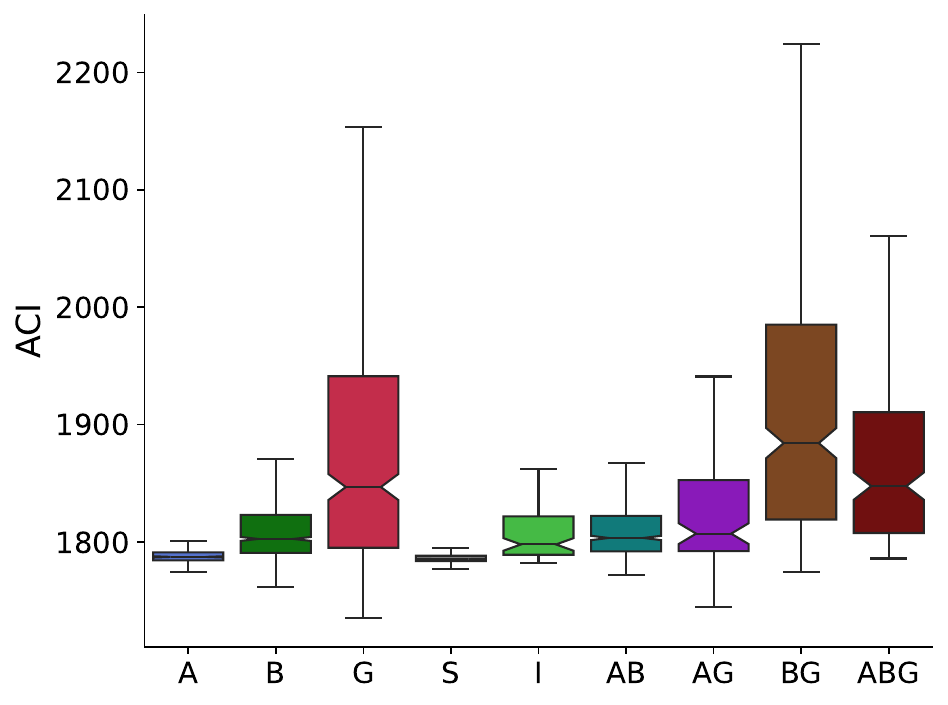}~%
    \includegraphics[width=.33\textwidth]{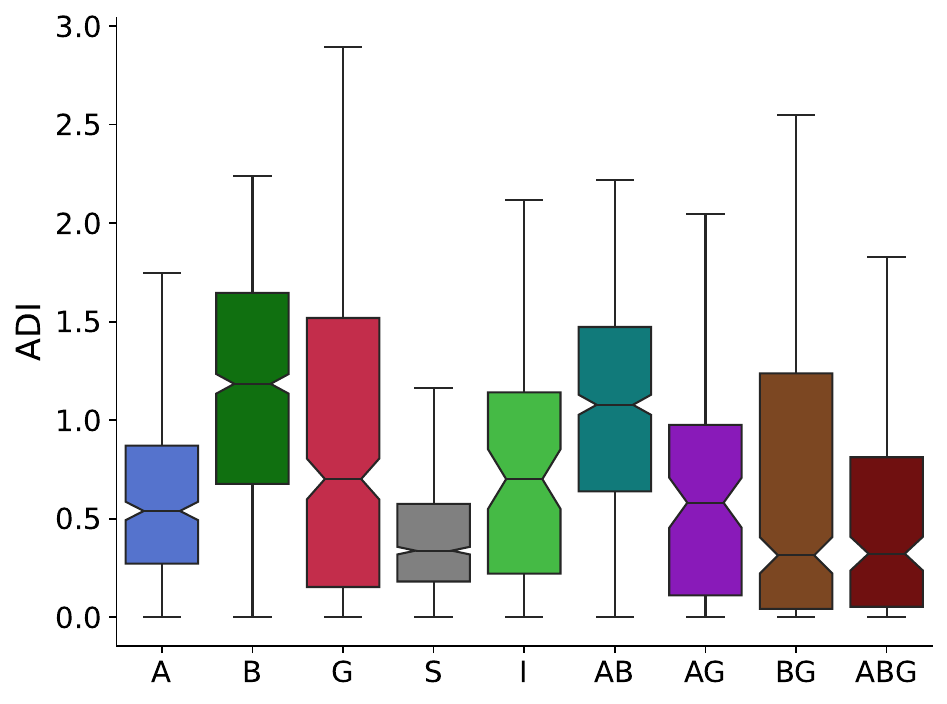}~%
    \includegraphics[width=.33\textwidth]{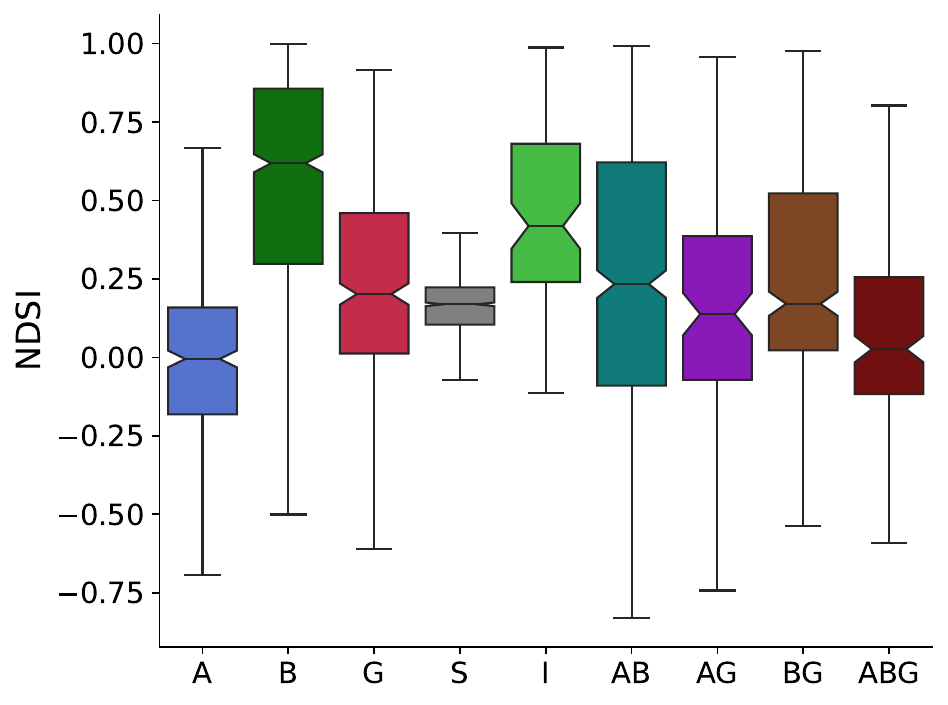}

    \includegraphics[width=.33\textwidth]{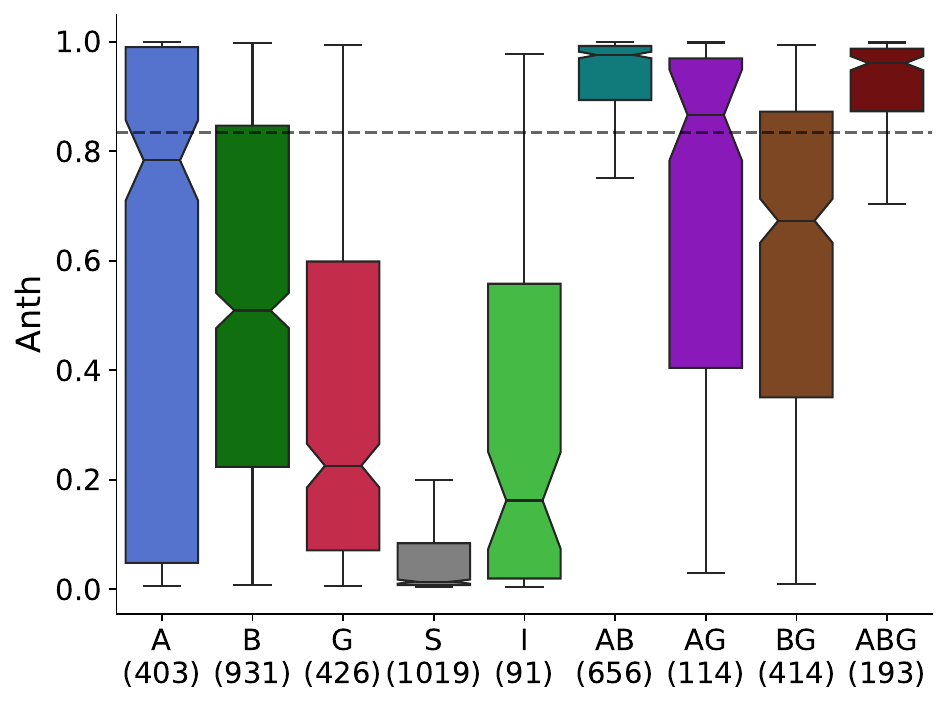}~%
    \includegraphics[width=.33\textwidth]{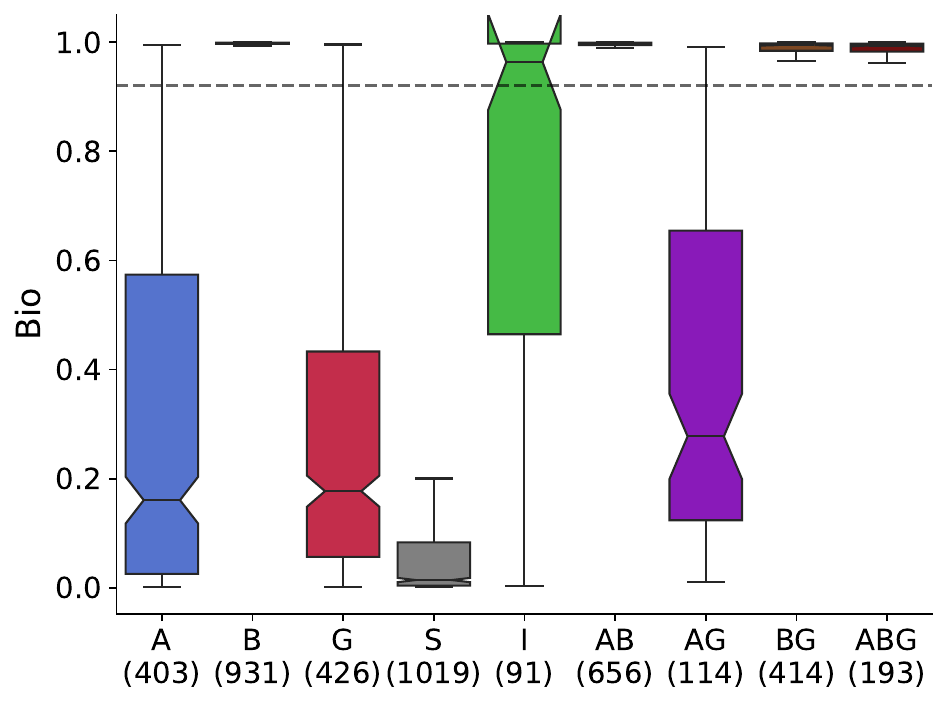}~%
    \includegraphics[width=.33\textwidth]{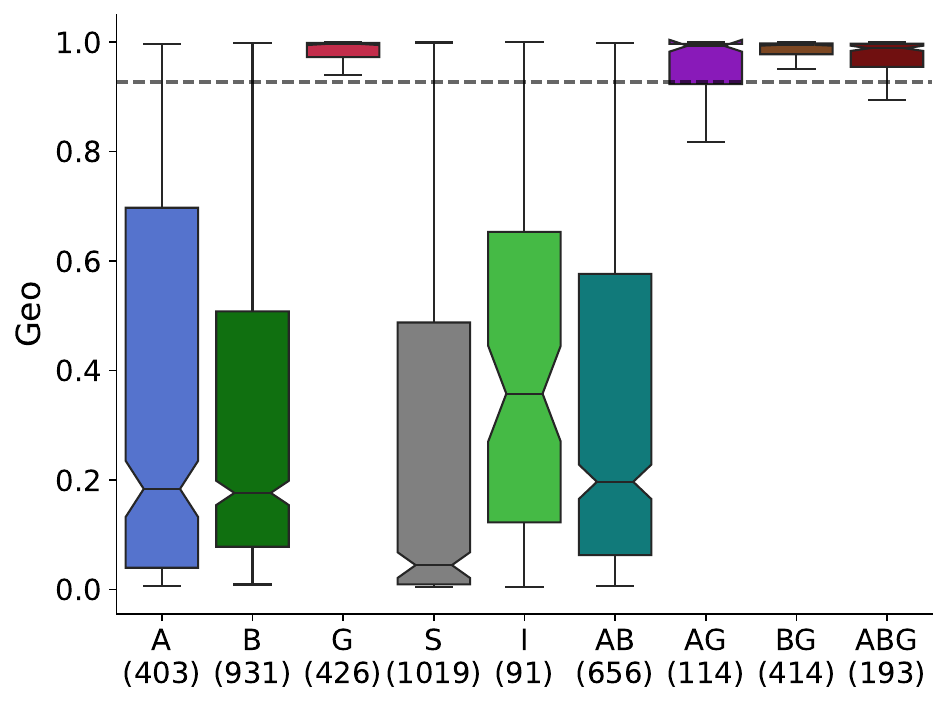}
    \caption{
    Distribution boxplots for ecoacoustic indices (top) vs CoarseSoundNet model predictions on {\besound} data (bottom) stratified per different label combinations:
    \emph{A}/\emph{B}/\emph{G}/\emph{S} denotes files only labelled as anthropophony/biophony/geophony/silence;
    \emph{I} is a subclass of biophony and denotes files labelled with insect sounds without the presence of anthropophony or geophony;
    \emph{A+B} denotes files labelled both with anthropophony and biophony;
    \emph{A+G} denotes files labelled both with anthropophony and geophony;
    \emph{B+G} denotes files labelled both with biophony and geophony;
    \emph{A+B+G} denotes files labelled with anthropophony, biophony, and geophony. Sample sizes for each label combination are given below the respective categories (based on the applied \ac{PDA}), and the class-specific decision thresholds (\ac{CST}) utilised are visualised as horizontal dashed lines.
    }
    \label{fig:coarse-vs-indices}
\end{figure}

\cref{fig:coarse-vs-indices} presents the distributions of three standard ecoacoustic indices (\ac{ACI}, \ac{ADI}, \ac{NDSI}; top row) and {\coarsenet} model predictions (bottom row) on the {\besound} data.
We have grouped predictions according to the underlying labels (on the file-level) by considering label combinations. 
We note that {\coarsenet} shows high reliability in its predictions for the respective class even in the presence of other classes.

For the ecoacoustic indices, a high overlap between classes is observed. \ac{ACI} values are lowest and least variable for segments labelled as silence or containing a single sound class, while higher median values and wider distributions are observed when geophony is present; particularly in combinations involving biophony and geophony (BG, ABG). 
\ac{ADI} shows higher median values for biophony (B) and biophony together with anthropophony (AB), but also exhibits considerable overlap and spread across mixed-label conditions. \ac{NDSI} values tend to be higher for segments containing biophony and lower for anthropophony-only segments, while mixed-class segments again span a wide range of values, often overlapping strongly with single-class distributions. 
Even though the distinction between anthropophony and biophony seems to be reasonable, especially for \ac{ACI} and \ac{NDSI}, there is a huge overlap with the other classes and combinations.

In contrast, the {\coarsenet} outputs exhibit a stronger separation aligned with the annotated labels. Anthropophony prediction scores are high for audios labelled with anthropophony alone or in combination with other classes (AB, AG, ABG), but have quite a wide distribution below the median for anthropophony-only where it overlaps with all the other labels and label combinations.
Similarly, biophony prediction scores are highest for biophony-only and insect-only recordings, as well as for combinations including biophony, while remaining low when biophony is absent. However, the biophony prediction scores on the insect recordings have quite a wide distribution below the median, which leads to overlap with anthropophony
and AG.
Geophony predictions show high scores for geophony-only segments and for segments where geophony co-occurs with other sound classes, and low scores otherwise.

These findings suggest that {\coarsenet} provides class-specific predictions that more clearly distinguish the underlying soundscape components than the distributions of the considered ecoacoustic indices, particularly when multiple sound classes co-occur.

\subsection{Ecological case study}
\label{ssec:results-ecological-case-study}

\begin{figure}[t]
    \centering
    \includegraphics[width=\textwidth]{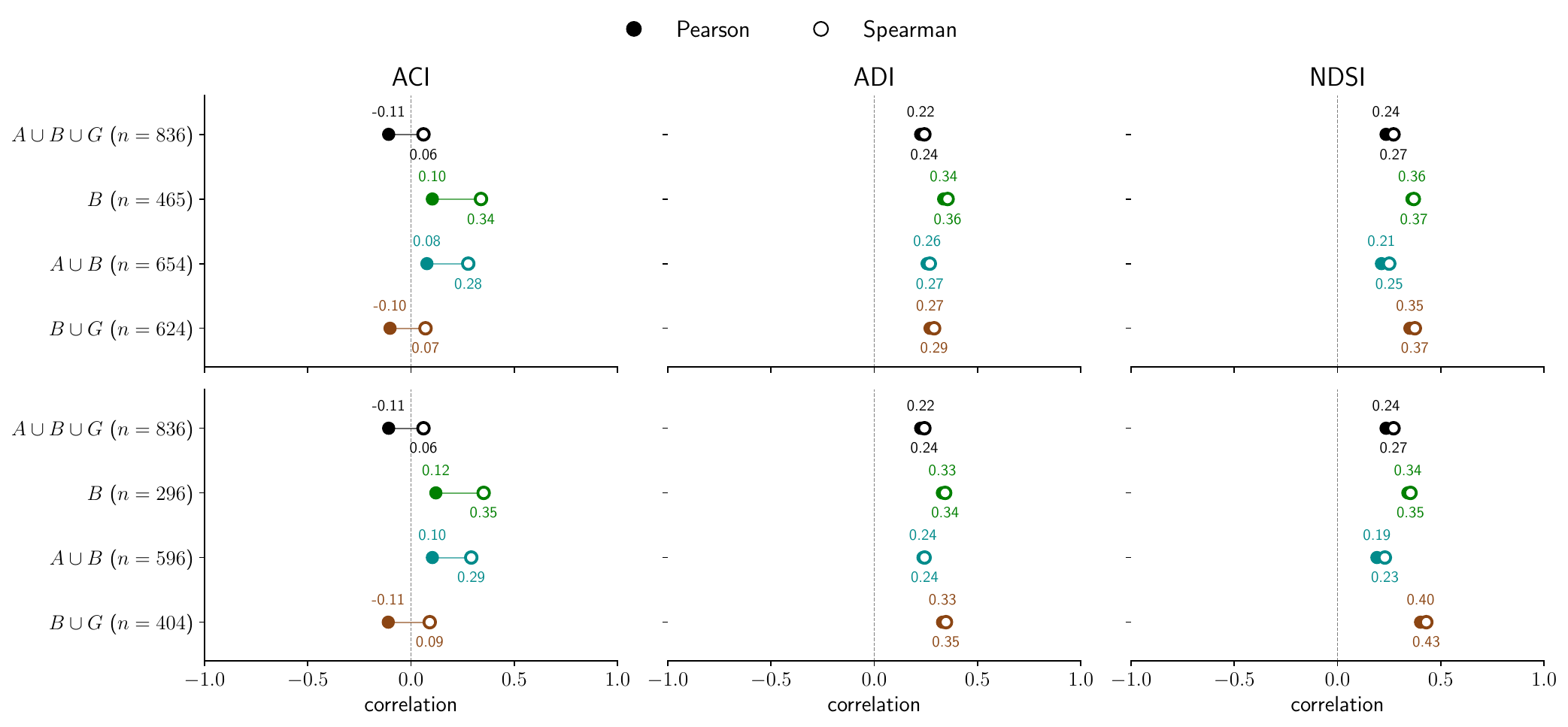}
    \caption{
    Pearson and Spearman correlations of three standard ecoacoustic indices (\ac{ACI}, \ac{ADI}, \ac{NDSI}) with $\alpha$-diversity (number of bird species identified from a human expert) for all data ($x \in A \cup B \cup G$) or filtered data using either the oracle values from human annotations (top row) or model predictions from our {\coarsenet} (bottom row).
    We filtered for:
    a) data containing only biophonic sounds ($x \in B$), green line;
    b) data containing both biophonic and anthropophonic sounds ($x \in A \cup B$), blue line;
    c) data containing both biophonic and geophonic sounds ($x \in B \cup G$), brown line. 
    }
    \label{fig:case-study-correlations}
\end{figure}

The results of the case study are visualised in \cref{fig:case-study-correlations,fig:case-study-regression}.
All indices are only weakly correlated with $\alpha$-diversity, with a best Pearson coefficient $r$ of only $.34$ and $.36$ obtained for \ac{ADI} and \ac{NDSI}, respectively, when only considering sounds without background noise and filtering using the human annotations.
These drop to $.22$ and $.24$, respectively, when considering all data.
Similar behaviour can be observed for the Spearman coefficient $\rho$.
However, for \ac{ACI} $r$ and $\rho$ diverge markedly, with $\rho$ showing a consistently more positive correlation than $r$, most notably for the biophony-only ($B$) and anthropophonic and biophonic ($A \cup B$) subsets.
For \ac{ADI} and \ac{NDSI}, filtering with {\coarsenet} instead of human annotations leaves $r$ and $\rho$ largely unchanged, except for the biophony and geophony subset ($x \in B \cup G$), where both coefficients increase (\ac{ADI}: $r$ from $.27$ to $.33$; \ac{NDSI}: $r$ from $.35$ to $.40$).
For \ac{ACI}, the already weak and partly negative correlations show no consistent pattern between the two filtering sources.

In general, filtering based on human annotations and {\coarsenet} predictions leads to very similar results.

\section{Discussion}
\label{sec:discussion}

\subsection{Deep learning architectures}
\label{ssec:discussion-models}
The performance differences observed between the {\edansa}-test and {\besound} in  \cref{ssec:results-model-training} highlight the strong influence of dataset characteristics on model generalisation. 
While models pretrained exclusively on ImageNet (\eg, ResNet-50 and EfficientNet-B7) or AudioSet (\eg, CNN10, CNN14, AST) achieve the best results on {\edansa}, models pretrained on large-scale and heterogenous audio dataset combinations (\eg, CLAP or Qwen2) perform most strongly on {\besound}. This suggests that broader and more diverse pretraining data can improve robustness to the noise and variability present in {\besound}. In addition, the substantially larger model capacities of these foundation models may enable the learning of more nuanced acoustic representations~\citep{Triantafyllopoulos25-CAF,Bommasani21-OTO}.

BirdNET, when used through the publicly available BirdNET Analyzer with default autotune, achieved lower performance than most of the fully fine-tuned models on both test sets. However, this comparison should be interpreted cautiously, as BirdNET was not fine-tuned under conditions directly comparable to the other architectures. The current interface of BirdNET does not finetune the pretrained model, but rather extracts embeddings from it and trains the final prediction layer.
Recent work has shown that a finetuning of all layers is necessary to obtain good downstream performance in transfer learning for audio tasks~\citep{Triantafyllopoulos21-TRO}. Furthermore, it utilises fixed $3\,s$ windows which might also be a limitation for our mostly $10\,s$ long training samples.
Indeed, the new version of BirdNET will enable these types of adaptations~\citep{Lasseck25-BVM}; however, it was not yet available to us.

The consistent performance drop observed across all models when transferring from {\edansa}-test to {\besound} suggests a substantial domain gap between the two datasets. This gap particularly affects anthropophony and geophony, which may be more sensitive to changes in recording conditions, background noise, and sound event prominence. In contrast, biophony appears to be more stable across domains, potentially due to its more distinctive acoustic patterns.

\subsection{The role of silence}
\label{ssec:discussion-silence}

We observe an improved model performance on {\besound} when including a silence class as a fourth category during training, as can be inferred from \cref{tab:besound-silence-results}. 
We assume that, without a silence class, the feature extractor is not sufficiently supervised to learn a representation of low-energy or non-event segments. 
Introducing a silence target anchors these segments to a dedicated region of the shared embedding space, preventing them from contaminating the representations of the three meaningful sound categories.
Thus, forcing the model to discriminate between meaningful and silent segments seems to encourage more robust feature learning, leading to better model performance.

Silence, in this context, does not necessarily refer to true acoustic silence. Rather, it represents segments without identifiable biophony, geophony, or anthropophony and can therefore also include recorder self-noise and low-energy, non-informative background sounds, similar to the concept of ambient sounds described by \citet{Grinfeder22-WDW}. Thus, silence serves primarily as a residual background class. Consequently, the acoustic characteristics of this class may vary across datasets depending on factors such as the annotation protocol, habitat, recording device, and gain settings.

\subsection{Impact of additional training data}
\label{ssec:discussion-data}

The results in \cref{tab:data-variations-edansa} show that adding additional data to {\edansa} does not improve the model performance on the {\edansa}-test set and in most cases even degrades it. In contrast, the results for {\besound} in \cref{tab:data-variations-besound} show a different trend, as the performance mostly improves, particularly when combining multiple external datasets. 
Adding a single dataset is beneficial only in some cases, specifically for the \ac{BE}-related datasets {\htsforest} and {\beambient}, arguably due to their greater domain similarity to {\besound}. 
Combining these two datasets already yields the second-best overall performance, while using all \ac{PAM} datasets together achieves the best results.
This is likely driven by the increased acoustic variability covered across datasets, including
differences in recording conditions such as microphone characteristics and geographic regions. Although each \ac{PAM} dataset already represents real-world soundscapes, it still has domain-specific biases. Therefore, combining multiple \ac{PAM} datasets reduces this mismatch and leads to more robust representations.

Finally, we observe that adding the mixed data (\synthdata) leads to the weakest results overall, performing even worse than the baseline. Despite careful design to make them realistic, these sound segments still do not seem to capture the full acoustic variability and complexity of real-world soundscapes. This contrasts previous bioacoustic studies, where adding synthesised data has improved model performance \citep{Guei24-EBS,Hoffman25-SDE,Gibbons24-GAI,Soltero25-RBD}. 

One possible explanation is that our mixing approach might introduce artefacts or unrealistic overlaps, potentially leading to classes masking each other too much. An option to mitigate this issue could be utilising silent \ac{PAM} recordings as a ``clean'' background onto which sound events are mixed, similar to \citet{Soltero25-RBD}. Furthermore, the relatively short duration of the samples ($5$\,s) may be a limiting factor as well.
In general, \citet{Eigenschink23-DGM} emphasise that realism and coherence are important factors for the usage of synthetic data in the audio domain. In light of this, and given the positive results of some prior bioacoustic studies, 
we argue that the use of synthetic data remains a promising direction for improving model performance, despite the limited gains observed in our experiments.

\subsection{Evaluation strategy and target-domain calibration}
\label{ssec:discussion-evaluation-strategy}
The results in \cref{ssec:results-evaluation-strategy} clearly show that applying class-specific confidence thresholds (\ac{CST}) benefits all three main classes. This is consistent with previous findings \citep{Scanferla25-DST,Arend25-SEO,Tseng25-SBC,Funosas26-AGA}, where class-dependent thresholding also improved performance. 
While biophony and anthropophony showed strong improvements, the effect on geophony was comparatively limited.

For biophony, we intentionally did not apply any time-based adjustments (\ac{PDA}) to the annotations. Many biophonic events, such as short bird calls, are naturally brief, and introducing a minimum-duration constraint would risk discarding valid detections. 
Since biophony tends to be easier to label reliably and less confusable with background noise than anthropophony or geophony, we focus primarily on \ac{CST} for this class. This leads to a noticeable relative performance increase of $3\%$. In contrast, combining \ac{CST} with the count-based method slightly degrades performance, indicating that additional temporal constraints are unsuitable for short, impulsive biophonic events.

For anthropophony, the first performance improvement is achieved by applying \ac{CST}, and this gain increases further when combining it with \ac{PDA}. Using \ac{PDA} alone together with a global threshold does not outperform the baseline, which indicates that threshold selection remains the decisive factor. The combination of \ac{CST} and \ac{PDA} yields the best results, suggesting that anthropogenic sounds benefit both from confidence calibration and from enforcing a minimum plausible event duration.

Geophony behaves differently as it improves more strongly when using solely \ac{PDA} than when using solely \ac{CST}. This might reflect the fact that geophonic sources (\eg, wind) are typically sustained over longer periods, making duration a natural indicator of reliability. As with anthropophony, combining \ac{CST} and \ac{PDA} leads to the strongest performance overall. The fact that the best-performing configuration uses a \ac{PDA} window of $15\,s$ for both anthropophony and geophony further supports the interpretation that these classes rely on longer and more temporally stable sound events.

When the count-based approach is added on top of \ac{PDA} and \ac{CST}, only geophony benefits further. This again suggests that geophonic events specifically profit from a broader temporal context, while especially biophony does not gain from this extra level of temporal smoothing. Biophony might have already reached the top end of possible performance with solely applying \ac{CST}.

\subsection{Error analysis and annotation quality}
\label{ssec:discussion-error-analysis}

The error analysis in \cref{ssec:results-error-analysis} indicates that label interactions are a primary source of the model confusions, with errors increasing notably when multiple sound classes co-occur. In particular, the presence of acoustically dominant sound types appears to mask quieter or less salient events, reducing their detectability. This effect is most pronounced for geophony and anthropophony, which show increased FN rates in mixed-class segments, whereas biophony remains comparatively robust. Since anthropophony has the highest FP and FN rates, this might indicate that this is a more difficult class per se. Specifically, events, such as far-off traffic, distant airplane sounds, or light footsteps, can be quite subtle and easy to miss or suppressed by other more dominant sounds.   
This is especially important for annotators, as they need to be aware and pay attention to the subtleties of the soundscapes, depending on how fine-grained and accurate the annotations shall be.

In this context, we further investigated annotation quality by randomly sampling and reviewing $1200$ recordings from the {\besound} data. The review was conducted by three of the authors, with each recording assessed by exactly one reviewer. For every recording, the presence of the three main classes, as well as silence, was re-evaluated and compared against the original annotations \wrt the whole $60$\,s.
This lead to the following mismatch percentages: $9.4\,\%$ for anthropophony, $1.5\,\%$ for biophony, $7.4\,\%$ for geophony, and $4.5\,\%$ for silence. 

These results reflect the varying difficulty of annotating each class. Biophony appears to be the most consistently and reliably annotated category, whereas geophony and especially anthropophony exhibit higher mismatch rates, indicating greater annotation difficulty.
These observations suggest that annotation difficulty is, to some extent, aligned with model performance across classes. Thus, unavoidable annotation noise likely introduces bias into the training data, which may further impact model performance.

Another notable observation is that the majority of biophony FNs correspond to insect sounds. This can partly be attributed to their limited representation in the training data, but may also result from the acoustic characteristics of certain insects, which stridulate predominantly at higher frequency ranges. 
Consequently, relevant spectral patterns of these signals are either strongly attenuated or entirely absent in the extracted features, and thus in {\coarsenet}'s input. 
Capturing such signals more reliably might therefore require increasing the audio sampling rate. 

Example spectrograms illustrating some of the main failure modes are provided in \cref{fig:error-analysis-spectrograms}.

\subsection{CoarseSoundNet vs Ecoacoustic Indices}
\label{ssec:discussion-coarsenet-vs-indices}

The comparison between standard ecoacoustic indices and {\coarsenet} predictions
in \cref{ssec:results-coarsenet-vs-indices} highlights fundamental differences in their ability to resolve complex soundscapes. While the indices capture broad class-dependent trends, especially when only focusing on biophony and anthropophony, their distributions show strong overlap across classes and label combinations, particularly in mixed-class scenarios, which limits their discriminative power.

In contrast, {\coarsenet} produces label-consistent prediction distributions even in the presence of multiple co-occurring sound classes, supporting its robustness to acoustic interference and masking. Nevertheless, both approaches show reduced reliability for insect sounds, where increased variability and overlap persist. This suggests that insects remain a challenging acoustic category, likely due to their spectral characteristics due to higher frequencies, and underscores a shared limitation of both index-based and learning-based methods in this domain. The wide distribution of the anthropophony predictions of {\coarsenet} again shows that this target class remains challenging. This difficulty is substantiated by the annotation errors of human experts when annotating anthropophony, as discussed in \cref{ssec:discussion-error-analysis}. 

Overall, the {\coarsenet} error patterns identified in \cref{ssec:results-error-analysis,ssec:discussion-error-analysis} are further confirmed by the prediction distributions presented in the boxplots.

\subsection{Ecological case study}
\label{ssec:discussion-ecological-case-study}

The results of the case study presented in \cref{ssec:results-ecological-case-study} show that the considered ecoacoustic indices are only weakly associated with avian $\alpha$-diversity in the {\besound} recordings.
The indices which had the highest correlation with $\alpha$-diversity are the \ac{ADI} and \ac{NDSI}, which is in line with previous studies in temperate regions, where the \ac{NDSI} could also achieve high correlations \citep{Eldridge18-SOE,Shaw24-FSH,Bradfer20-RAO}. 
The \ac{ADI} was even more correlated with species richness by \citet{Eldridge18-SOE} than the \ac{NDSI}, while there was no significant correlation reported by \citet{Shaw24-FSH}. In our case, both indices perform almost on par, especially when filtering out everything else than biophony. 

However, filtering background noise and non-biophonic sound sources using {\coarsenet}, as well as filtering based on human-annotated ground truth, yielded at best 
marginal improvements. This indicates that, when applied in isolation, these indices are not sufficient to reliably capture ecological complexity in acoustically heterogeneous environments. 
In contrast, \citet{Jiang26-RNS} achieved higher correlations between their avian sound class and acoustic indices after removing anthropophonic, geophonic, and insect sounds in an urban environment. 
Together with the findings reported in \cref{ssec:results-coarsenet-vs-indices,ssec:discussion-coarsenet-vs-indices}, this suggests that standard ecoacoustic indices, while useful as coarse proxies, benefit from complementary approaches, such as those provided by {\coarsenet} or other eco- and bioacoustic \ac{ML} models, to more robustly characterise biodiversity patterns in \ac{PAM} soundscapes.

Specifically, quantifying the soundscape components is not only a tool to improve acoustic index performance, but also a valuable approach to test ecoacoustic hypotheses and improve interpretability of acoustic patterns. Moreover, being able to attribute changes in an acoustic index to, \eg, an increase in biophony, and not geophony or anthropophony, strengthens the interpretation of the results.

\subsection{Limitations and external validity}
\label{ssec:external-validity}
Since the model was trained solely on terrestrial \ac{PAM} datasets, with strong representation from Arctic Alaska and temperate habitats in Germany, its performance may not directly transfer to acoustically distinct environments such as tropical forests, wetlands, urban soundscapes, or marine ecosystems. Furthermore, as the final model operates at a sampling frequency of $48$\,kHz, frequencies above the corresponding Nyquist limit are not represented, and the model is therefore not designed to capture ultrasonic taxa. In addition, differences in recording hardware and microphone frequency responses may further affect model performance. Future work should evaluate the generalisability of {\coarsenet} across a broader range of ecosystems, recording devices, and acoustic conditions.

A further limitation concerns the dual role of {\besound} in model development and evaluation. While our broader objective was to develop a model that transfers across different terrestrial \ac{PAM} settings, {\besound} represents a particularly relevant target domain and its performance was deliberately considered when selecting several model configurations. Its annotations were additionally used for post-hoc target-domain calibration. Consequently, the final performance reported on {\besound} reflects the deliberately calibrated performance for this target domain and should be interpreted accordingly, rather than an estimate of generalisation to entirely unseen domains.

\section{Conclusion}
In this study, we trained the {\coarsenet} model in a multi-label setting in order to distinguish between the three coarse soundscape classes anthropophony, biophony, and geophony. We first selected a suitable model architecture and then explored several model optimisation approaches, including adding an additional ``silence'' class, as well as the integration of various additional training data. Subsequently, we examined different evaluation strategies, such as class-specific confidence thresholds and time-based annotation adjustments. 
We then conducted an error analysis of the model, compared its outputs to three classical acoustic indices, and finally illustrated a potential ecological application in a case study.

Our findings show that adding more training data, especially from domains that closely match the target data conditions, further boosts the model performance.
Furthermore, adding an additional silence class during training improved the discrimination between the three main soundscape components. 
Regarding the evaluation strategies, we recommend using class-specific thresholds if possible, as they consistently improve performance across all three classes.
For anthropophony and particularly for geophony, the additional use of duration-based constraints yields even more performance gains, reflecting the typically longer and more temporally continuous nature of these signals.  

The error analysis indicates that anthropophony is particularly challenging, as it might be masked or suppressed by biophonic or geophonic sounds, thus requiring especially careful annotation. For geophony, silence is the most pronounced source of confusion, while insects generally pose a source of error, leading in particular to false negatives for biophony. 

Finally, the ecological case study demonstrated that filtering the recordings with {\coarsenet} before computing ecoacustic indices can yield similar trends to those using ground-truth filtering. However, the correlation between the indices and $\alpha$-diversity remains generally rather weak.
Thus, while {\coarsenet} enabled automated soundscape-based filtering that produced downstream trends broadly similar to those obtained with ground-truth filtering, filtering provided only limited benefits for relating the considered acoustic indices to avian $\alpha$-diversity.

\section*{CRediT authorship contribution statement}
\textbf{Schuller Björn W.:} Writing – review \& editing, Supervision, Resources, Project administration, Investigation, Funding acquisition, Conceptualisation. \textbf{Michael Scherer-Lorenzen:} Writing – review \& editing, Supervision, Resources, Project administration, Funding acquisition. 
\textbf{Svenja Schmidt:} Writing – review \& editing, Data curation.
\textbf{Sandra Müller:} Writing – review \& editing, Validation, Supervision, Project administration, Methodology, Investigation, Funding acquisition, Formal analysis, Data curation, Conceptualisation. \textbf{Dominik Arend:} Writing – review \& editing, Validation, Resources, Methodology, Investigation, Formal analysis, Data curation. \textbf{Andreas Triantafyllopoulos:} Writing – review \& editing, Validation, Resources, Project administration, Methodology, Software, Investigation, Funding acquisition, Formal analysis, Conceptualisation. \textbf{Alexander Gebhard:} Writing – original draft, Validation, Resources, Methodology, Software, Investigation, Formal analysis, Data curation, Conceptualisation.

\section*{Declaration of Competing Interest}
The authors declare that they have no known competing financial interests or personal relationships that could have appeared to influence the work reported in this paper.

\section*{Acknowledgements}
\label{sec:acknowledgements}
We thank the managers of the three Exploratories, Julia Bass, Max Müller, Anna K. Franke, Robert Künast, Franca Marian, Melissa Jüds and all former managers for their work in maintaining the plot and project infrastructure; Victoria Griessmeier for giving support through the central office, Andreas Ostrowski for managing the central data base, and Markus Fischer, Eduard Linsenmair, Dominik Hessenmöller, Daniel Prati, Ingo Schöning, François Buscot, Ernst-Detlef Schulze, Wolfgang W. Weisser and the late Elisabeth Kalko for their role in setting up the Biodiversity Exploratories project. We thank the administration of the Hainich national park, the UNESCO Biosphere Reserve Swabian Alb and the UNESCO Biosphere Reserve Schorfheide-Chorin as well as all land owners for the excellent collaboration. 
We also thank Robert Künast for reading the manuscript and providing valuable feedback.
The work has been (partly) funded by the DFG Priority Program 1374 "Biodiversity-Exploratories" (512414116). Field work permits were issued by the responsible state
environmental offices of Baden-Württemberg, Thüringen, and Brandenburg.

\section*{Data Availability}
\label{sec:data-availability}

\begin{itemize}
    \item The {\coarsenet} model is available via huggingface: \url{https://huggingface.co/HearTheSpecies/CoarseSoundNet}.  
    \item The corresponding code is available on github:  \url{https://github.com/CHI-TUM/CoarseSoundNet}.
    \item The {\edansa} data is publicly available \citep{Coban22-ETE}.
    \item The \akwamo, \htsforest, \beambient, \synthdata, and {\besound} data are accessible on zenodo: \url{https://zenodo.org/records/21907058}.
\end{itemize}

\appendix
\section{Supplementary material}
\label{app:supp-material}

\subsection{Data}
\label{app:data}
\Cref{tab:data-roles} gives an overview of the roles of the datasets across the experimental pipeline. {\edansa} was used for model training in every experimental setup while its corresponding held-out test set was then used for evaluation model performance and inform model-development decisions. The additional \ac{PAM} datasets were leveraged for training only during the training-data composition experiments, whereas {\besound} was not utilised for model training but served as a target-domain dataset for evaluation, model-development decisions, calibration, and subsequent analyses.

\begin{table}[ht]
\centering
\caption{Overview of the roles of the datasets across the experimental pipeline. A checkmark indicates that a dataset was used in the respective experimental stage; depending on the stage, datasets were used for model training or for evaluating and guiding model-development decisions. Asterisks indicate additional \ac{PAM} datasets that were used for training only during the training-data composition experiments. {\besound} was not used for model training.
}
\resizebox{\textwidth}{!}{%
\begin{tabular}{lccccccc}
\hline
\textbf{Dataset} &
\makecell{\textbf{Model}\\\textbf{training}} &
\makecell{\textbf{Architecture}\\\textbf{selection}} &
\makecell{\textbf{Silence-class}\\\textbf{selection}} &
\makecell{\textbf{Training-data}\\\textbf{composition}} &
\makecell{\textbf{Target-domain}\\\textbf{calibration}} &
\makecell{\textbf{Error}\\\textbf{analysis}} &
\makecell{\textbf{Ecological}\\\textbf{case study}} \\
\hline
{\edansa}  & \checkmark & \checkmark & \checkmark & \checkmark & -- & -- & -- \\
{\akwamo}  & \checkmark$^{*}$ & -- & -- & \checkmark & -- & -- & -- \\
{\beambient} & \checkmark$^{*}$ & -- & -- & \checkmark & -- & -- & -- \\
{\htsforest} & \checkmark$^{*}$ & -- & -- & \checkmark & -- & -- & -- \\
{\synthdata}  & \checkmark$^{*}$ & -- & -- & \checkmark & -- & -- & -- \\
{\besound} & -- & \checkmark & \checkmark & \checkmark & \checkmark & \checkmark & \checkmark \\
\hline
\end{tabular}%
}
\label{tab:data-roles}
\end{table}

\subsection{Grid search parameters}
\label{app:grid-search}

\subsubsection{Different models}
\label{sssec:gs-different-models}
For the grid search, we explore the parameters listed in \cref{tab:gs-hparams}. The BirdNet grid search was conducted using the default autotune settings \url{https://birdnet-team.github.io/BirdNET-Analyzer}.

\begin{table}[t]
\centering
\caption{Training hyperparameters used in the grid-search of the models. The utilised batch sizes, learning rates, optimisers, and model variants are reported for each model.}
\label{tab:gs-hparams}
\resizebox{\textwidth}{!}{%
\begin{tabular}{l|c|c|c|c}
\textbf{Model} & \textbf{Batch Size} & \textbf{Learning Rate} & \textbf{Optimiser} & \textbf{Variant} \\
\hline
CNN 10 & $16$, $32$ & $.001$, $.0001$ & Adam, AdamW & -- \\
CNN 14 & $16$, $32$ & $.001$, $.0001$ & Adam, AdamW & -- \\
ResNet-50 & $16$, $32$ & $.001$, $.0001$ & Adam, AdamW & -- \\
EfficientNet-B7 & $16$, $32$ & $.001$, $.0001$ & Adam, AdamW & B7 \\
BirdNET & -- & -- & -- & -- \\
AST & $16$ & $.001$, $.0001$, $.00001$ & Adam, AdamW & -- \\
SSAST & $16$ & $.001$, $.0001$, $.00001$ & Adam, AdamW & -- \\
PaSST & $16$ & $.001$, $.0001$, $.00001$ & Adam, AdamW & -- \\
AVES & $16$ & $.001$, $.0001$, $.00001$ & Adam, AdamW & -- \\
W2V2 & $16$ & $.001$, $.0001$, $.00001$ & Adam, AdamW & base, large \\
Whisper & $16$ & $.001$, $.0001$, $.00001$ & Adam, AdamW & small \\ 
CLAP & $16$ & $.001$, $.0001$, $.00001$ & Adam, AdamW & -- \\
Qwen2-Audio & $4$ & $.001$, $.0001$, $.00001$ & Adam, AdamW & -- \\
\end{tabular}
}
\end{table}

\subsubsection{The role of silence}
\label{sssec:gs-silence}

The parameters considered in the grid search of the chosen models for the silence experiment are listed in \cref{tab:gs-silence-hparams}.

\begin{table}[t]
\centering
\caption{Training hyperparameters used in the grid-search of the selected models. The batch sizes, learning rates, optimisers, and augmentation variants are reported for each model.}
\label{tab:gs-silence-hparams}
\resizebox{\textwidth}{!}{%
\begin{tabular}{l|c|c|c|c}
\textbf{Model} & \textbf{Batch Size} & \textbf{Learning Rate} & \textbf{Optimiser} & \textbf{Augmentation} \\
\hline
CNN 10 & $16$, $32$ & $.001$, $.0001$ & Adam & None, SpecAugment, Custom \\
AST & $16$ & $.0005$, $.0001$, $.00001$ & Adam & None, SpecAugment, Custom \\
CLAP & $16$ & $.0005$, $.0001$, $.00001$ & Adam & None, SpecAugment, Custom \\
\end{tabular}
}
\end{table}

\subsection{Model soups}
\label{app1}

\Cref{tab:appendix-model-soup} shows the macro F1-score for each \ac{MS}, averaging all the weights of each grid search run.

\begin{table}[t]
\centering
\caption{F1-Scores (macro) for each model soup evaluated on the {\edansa} test set. The \textit{soup} for each model refers to the average F1-score (macro) over all models trained during the grid search.}
\label{tab:appendix-model-soup}
\resizebox{\textwidth}{!}{%
\begin{tabular}{l|cccccccccccc}
    \textbf{Model} & \textbf{CNN 10} & \textbf{CNN 14} & \textbf{ResNet-50} & \textbf{EffNet-B7} & \textbf{AST} & \textbf{SSAST} & \textbf{PaSST} & \textbf{AVES} & \textbf{W2V2} & \textbf{Whisper} & \textbf{CLAP} & \textbf{Qwen2-Audio} \\
    \hline
    MS $\uparrow$ & .902 & .906 & .679 & .246 & .295 & .506 & .507 & .000 & .262 & .442 & .383 & .369 \\
\end{tabular}
}
\end{table}

\subsection{Additional spectrogram examples}
\label{app:spectrograms}

\begin{figure}[t]
    \centering
    \includegraphics[width=\textwidth]{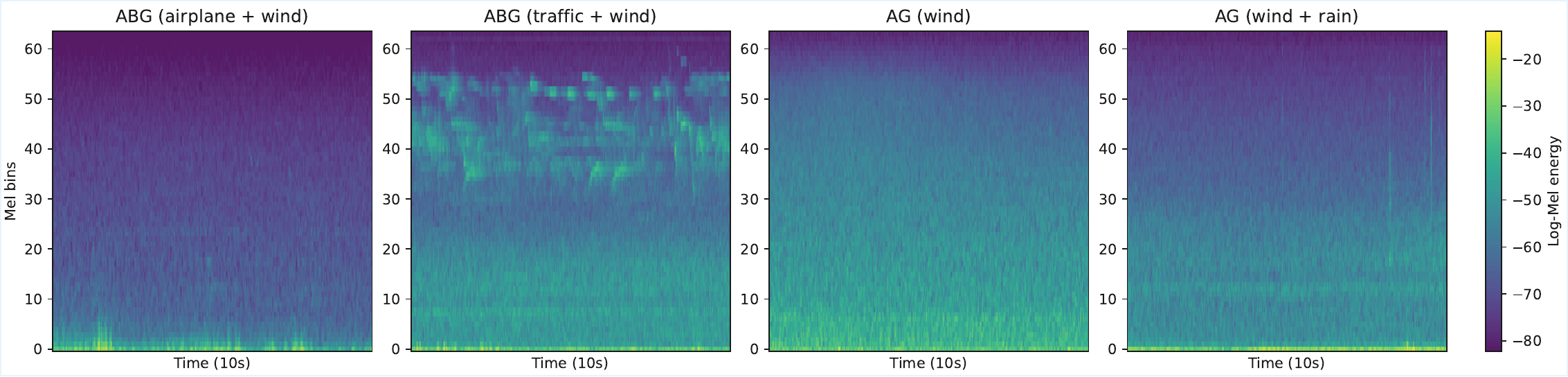}
    \caption{Example spectrograms of recordings with all three classes present (first two) as well as recordings with only anthropophony and geophony present (last two).}
    \label{fig:mixture-spectrograms}
\end{figure}

\cref{fig:mixture-spectrograms} illustrates examples of recordings in which multiple sound classes occur simultaneously, with a particular focus on anthropophony and geophony. The first two spectrograms depict instances where all three classes are present, whereas the final two contain only anthropophony and geophony. In the first spectrogram, bird calls are faint and distant, while in the second they are noticeably closer and louder, which is clearly reflected in the spectrograms. The anthropophonic component in the first example is an airplane, whereas traffic noise is present in the remaining three. In the final spectrogram, rain can also be observed in addition to wind.
These examples highlight the complexity of accurately annotating overlapping sound sources and demonstrate that achieving error-free annotations is a highly challenging task.

\begin{figure}[t]
    \centering
    \includegraphics[width=\textwidth]{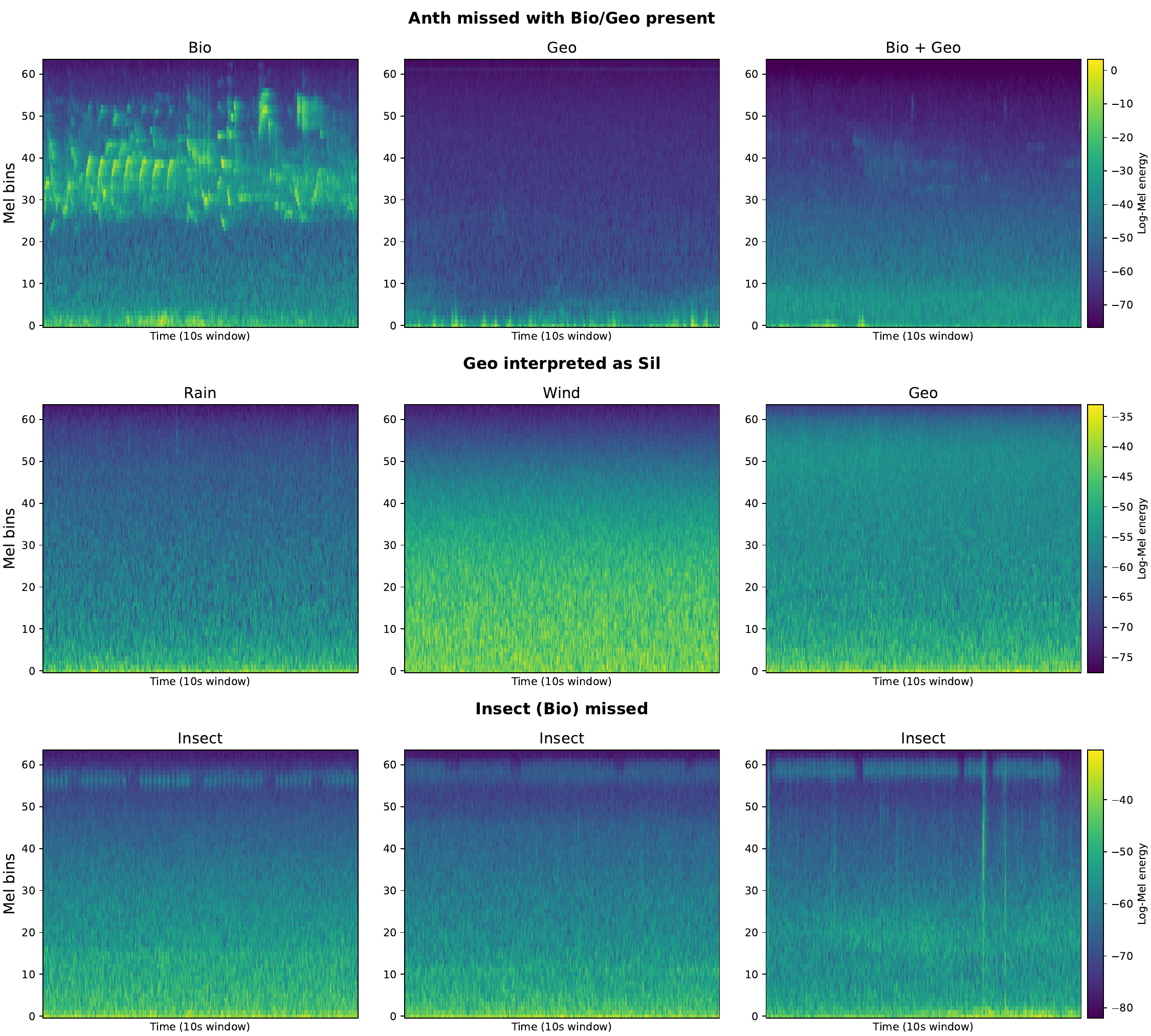}
    \caption{Example spectrograms illustrating the main failure modes identified in the {\coarsenet} error analysis on {\besound} after final threshold calibration. The top row shows examples of annotated anthropophony which has not been recognised in the presence of biophony and/or geophony. The middle row displays annotated geophony which was not classified but rather interpreted as silence. The bottom row presents examples of missed biophony corresponding to insect sounds.}
    \label{fig:error-analysis-spectrograms}
\end{figure}

\Cref{fig:error-analysis-spectrograms} illustrates some example spectrograms of the main failure modes identified in \cref{ssec:discussion-error-analysis}.

\subsection{Evaluation strategies}
\label{app:evaluation-strategies}
\Cref{tab:pda-counts} shows the number of positive annotations per target class in {\besound} for each \ac{PDA} threshold.

To provide a more rigorous evaluation of the post-hoc calibration strategies from \cref{ssec:evaluation-strategy}, we additionally performed three-fold cross-validation on {\besound}, with recordings grouped by recording plot. For each fold, the threshold parameters are selected using two folds and evaluated on the remaining fold. We report the mean and standard deviation of the resulting F1-scores across the three held-out folds in \cref{tab:threshold_variants_cv}.
This procedure separates calibration-parameter selection from evaluation. However, {\besound} had already informed earlier model-development decisions, including architecture selection, silence-class inclusion, and training-data composition. Consequently, these results evaluate post-hoc calibration conditional on those prior decisions and do not constitute an independent validation of the complete model-development pipeline.

\begin{table}[t]
\centering
\caption{Number of positive annotations per target class in {\besound} before (baseline) and after applying \ac{PDA} with minimum annotation durations of $p=5$ ($3\,s$), $p=10$ ($6\,s$), and $p=25$ ($15\,s$) of the $60\,s$ recording duration.}
\begin{tabular}{lrrr}
\hline
\textbf{Constraint} & \textbf{Anth} & \textbf{Bio} & \textbf{Geo} \\
\hline
baseline         & 1631 & 2285 & 1738 \\
$p=5$ ($3\,s$)   & 1468 & 2197 & 1376 \\
$p=10$ ($6\,s$)  & 1436 & 2132 & 1298 \\
$p=25$ ($15\,s$) & 1366 & 2023 & 1147 \\
\hline
\end{tabular}
\label{tab:pda-counts}
\end{table}

\begin{table}[t]
    \centering
    \caption{Recording-level multi-label classification performance for the post-hoc calibration strategies described in \cref{ssec:evaluation-strategy}, evaluated using three-fold cross-validation on {\besound} with folds grouped by recording plot. Values represent the mean F1-score $\pm$ standard deviation across the three folds.}
    \label{tab:threshold_variants_cv}
    \begin{tabular}{l|ccc|c}
        \hline
        Threshold variant & Anth & Bio & Geo & Macro $\uparrow$ \\
        \hline
        \ac{PDA}  & $.649 \pm .012$ & $.909 \pm .003$ & $.738 \pm .019$ & .766 $\pm .006$ \\
        \ac{CST}  & $.652 \pm .010$ & $.936 \pm .002$ & $.708 \pm .024$ & .765 $\pm .006$ \\
        \ac{PDA} + \ac{CST} & $.655 \pm .012$ & $.934 \pm .002$ & $.757 \pm .014$ & \textbf{.782} $\pm .003$ \\
        \hline
    \end{tabular}
\end{table}

\subsection{Final {\coarsenet} configuration \\and {\besound} calibrations}
\label{app:final-config}
\Cref{tab:final-configuration} lists the final model configurations and the recommended thresholds calibrated on {\besound}.
For post-hoc calibration to a specific target domain, we generally recommend \ac{CST} when representative annotated data are available. If the data is strongly annotated and thus additionally allows for the exclusion of short-duration annotations, \ac{CST} can further be combined with \ac{PDA}. However, \ac{CST} should then be jointly adjusted with \ac{PDA}.

\begin{table}[t]
\centering
\caption{Configuration of the final {\coarsenet} model and recommended post-hoc configurations for the {\besound} target domain. We report both recommended calibration strategies: \ac{CST} alone as well as \ac{PDA} + \ac{CST}.}
\begin{tabular}{lcc}
\hline
\textbf{Component} & \multicolumn{2}{c}{\textbf{Configuration}} \\
\hline
\multicolumn{3}{l}{\textit{Model configuration}} \\
Architecture      & \multicolumn{2}{c}{CLAP-HTSAST} \\
Training datasets & \multicolumn{2}{c}{{\edansa} + all \ac{PAM} datasets} \\
Silence class     & \multicolumn{2}{c}{Included} \\
Window size       & \multicolumn{2}{c}{$10\,s$} \\
Step size         & \multicolumn{2}{c}{$10\,s$} \\
\hline
\multicolumn{3}{l}{\textit{Post-hoc configuration for {\besound}:}} \\
\hline
\multicolumn{1}{l}{\textbf{Strategy}} & \multicolumn{2}{c}{\textbf{Threshold}} \\
\underline{\ac{CST}} & \multicolumn{2}{c}{\ac{CST}} \\
Anth     & \multicolumn{2}{c}{$.722$} \\
Bio      & \multicolumn{2}{c}{$.920$} \\
Geo      & \multicolumn{2}{c}{$.571$} \\
\hline
\underline{\ac{PDA} + \ac{CST}} & \ac{PDA} & \ac{CST} \\
Anth                & $15\,s$ & $.835$ \\
Bio                 & -- & $.920$ \\
Geo                 & $15\,s$ & $.927$ \\
\hline
\end{tabular}
\label{tab:final-configuration}
\end{table}

\subsection{Ecological case study}
\label{app:case-study}

\begin{figure}[t]
    \centering
    \includegraphics[width=\textwidth]{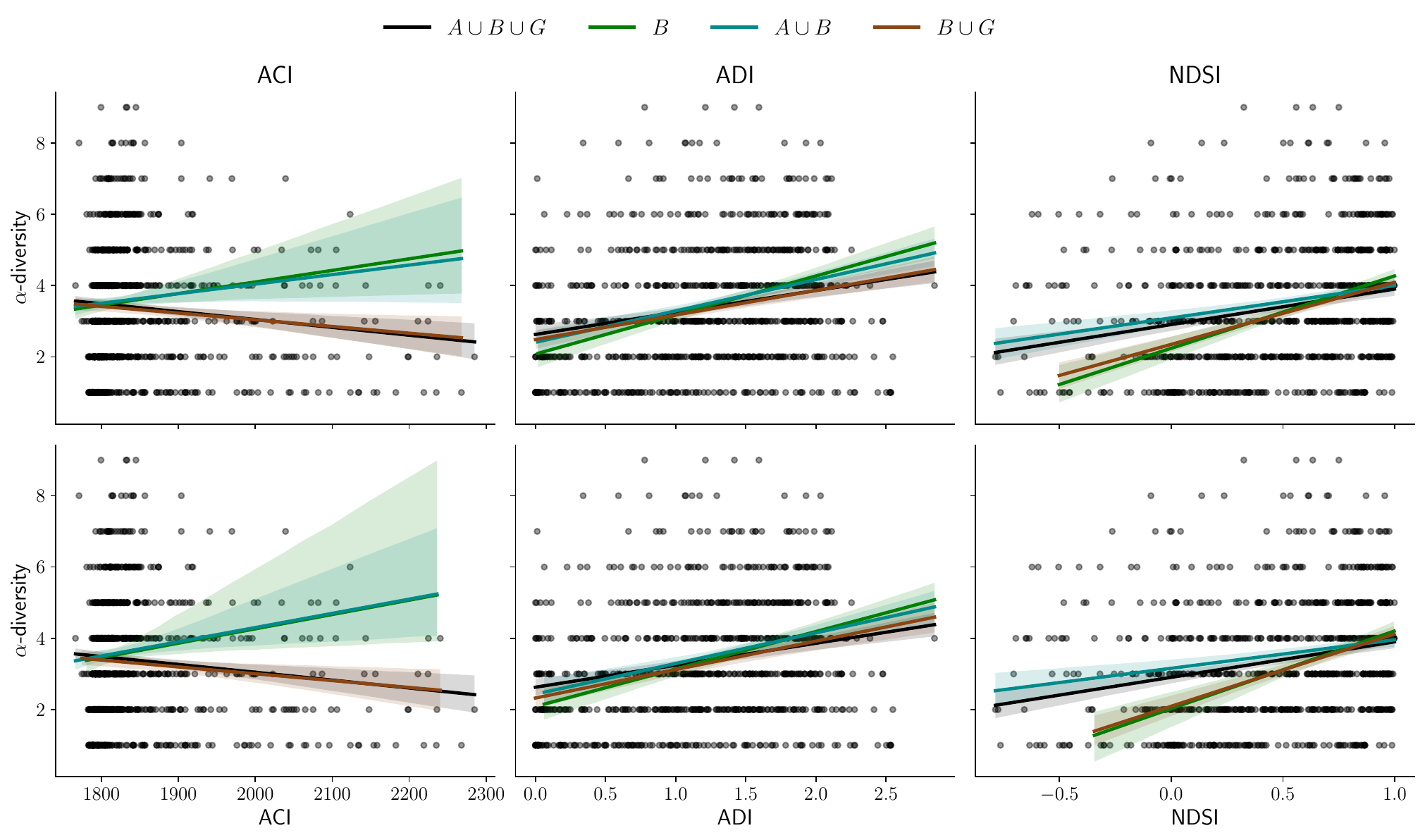}
    \caption{
    Regression of three standard ecoacoustic indices (\ac{ACI}, \ac{ADI}, \ac{NDSI}) with $\alpha$-diversity (number of bird species identified from a human expert) for all data ($x \in A \cup B \cup G$) or filtered data using either the oracle values from human annotations (top row) or model predictions from our {\coarsenet} (bottom row).
    We filtered for:
    a) data containing only biophonic sounds ($x \in B$), green line;
    b) data containing both biophonic and anthropophonic sounds ($x \in A \cup B$), blue line;
    c) data containing both biophonic and geophonic sounds ($x \in B \cup G$), brown line.
    }
    \label{fig:case-study-regression}
\end{figure}

\Cref{fig:case-study-regression} shows the scatter of three standard ecoacoustic indices (\ac{ACI}, \ac{ADI}, \ac{NDSI}) with $\alpha$-diversity (number of bird species identified from a human expert), together with regression fits for the same filtering conditions and colour coding as \cref{fig:case-study-correlations}: all data ($x \in A \cup B \cup G$), biophoniy-only data ($x \in B$), biophonic and anthropophonic sounds ($x \in A \cup B$), and data containing both biophonic and geophonic sounds ($x \in B \cup G$). The Pearson and Spearman correlation coefficients reported in \cref{fig:case-study-correlations} were computed from this same data.

\clearpage

\bibliographystyle{elsarticle-harv}
\bibliography{refs}
\end{document}